\documentclass[pra,reprint,aps,floatfix]{revtex4-2}
\usepackage{graphicx} % Include figure files
\usepackage{epstopdf}
\usepackage{dcolumn}% Align table columns on decimal point
\usepackage{bm}% bold math
\usepackage{amsfonts,amsmath,amssymb%,float
}

\usepackage{bm,dsfont}

\usepackage{bbm}
\usepackage{longtable}
\usepackage[dvips]{epsfig}
\usepackage{hyperref}
\usepackage{listings}
\usepackage{color}

\newcommand{\bra}[1]{\boldsymbol{\langle} #1\boldsymbol{|}}
\newcommand{\ket}[1]{\boldsymbol{|}#1 \boldsymbol{\rangle}}
\newcommand{\braket}[2]{\boldsymbol{\langle} #1\boldsymbol{|}#2 \boldsymbol{\rangle}}

\let\phi\varphi

\let\theta\vartheta
\renewcommand{\rho}{\varrho}
\newcommand{\uber}[2]{{\binom{#1}{#2}}}

\renewcommand{\vec}[1]{\ensuremath{\boldsymbol #1 }}

\begin{document}
\title{Dynamical generation of chiral $W$ and Greenberger-Horne-Zeilinger
states in laser-controlled Rydberg-atom trimers}
	
\author{Thorsten Haase, Gernot Alber, and Vladimir M. Stojanovi\'c}
\affiliation{Institut f\"{u}r Angewandte Physik, Technical
University of Darmstadt, D-64289 Darmstadt, Germany}
\date{\today}
\begin{abstract}
Motivated by the significantly improved scalability of optically-trapped neutral-atom systems, extensive efforts have been devoted 
in recent years to quantum-state engineering in Rydberg-atom ensembles. Here we investigate the problem of engineering generalized 
(``twisted'') $W$ states, as well as Greenberger-Horne-Zeilinger (GHZ) states, in the strongly-interacting regime of a neutral-atom 
system. We assume that each atom in the envisioned system initially resides in its ground state and is subject to several external 
laser pulses that are close to being resonant with the same internal atomic transition. In particular, in the special case of a 
three-atom system (Rydberg-atom trimer) we determine configurations of field alignments and atomic positions that enable the realization 
of {\em chiral $W$ states} -- a special type of twisted three-qubit $W$ states of interest for implementing noiseless-subsystem qubit 
encoding. Using chiral $W$ states as an example we also address the problem of 
deterministically converting twisted $W$ states into their GHZ counterparts 
in the same three-atom system, thus significantly generalizing recent works that involve only ordinary $W$ states. We show that starting from 
twisted -- rather than ordinary -- $W$ states is equivalent to renormalizing downwards the relevant Rabi frequencies. While this 
leads to somewhat longer state-conversion times, we also demonstrate that those times are at least two orders of magnitude shorter 
than typical lifetimes of relevant Rydberg states.
\end{abstract}
	
\maketitle
\section{Introduction}
Maximally-entangled multiqubit states are of special interest for quantum-information processing (QIP)~\cite{Nielsen:99}. 
Two particularly prominent classes of such states are $W$~\cite{Duer+:00} and GHZ~\cite{Greenberger+Horne+Zeilinger:89} states, 
for which it is known that they cannot be transformed into each other through local operations and classical communication (LOCC 
inequivalence~\cite{Nielsen:99}). Owing to their proven usefulness in various QIP protocols~\cite{Joo+:03,Agrawal+Pati:06,Zhu+:15,
Maleki+Zubairy:22}, a multitude of different schemes for the preparation of $W$~\cite{Tashima+:08,Li+Song:15,Kang+:16,Kang+SciRep:16,
StojanovicPRL:20,StojanovicPRA:21,Cole+:21,Peng+:21,Pachniak+Malinovskaya:21,Zheng++:22,Zhang+:22} and GHZ states~\cite{Song+:17,Erhard+:18,
Macri+:18,Zheng+:19,Nogueira+:21} in various physical platforms have been proposed in recent years. 
	
One of the currently most promising platforms for QIP is based on ensembles of neutral atoms in Rydberg states~\cite{GallagherBOOK,Adams+:20}.
The scalability of these systems, confined in arrays of individual optical-dipole microtraps (tweezers), has improved 
significantly in recent years~\cite{Barredo+:16,Bernien+:17,Barredo+:18,Brown+:19,deMello+:19,Schymik+:20}. This development
has been interwoven with other important feats, such as high-fidelity state preparation/readout and accurate realization of 
quantum logic gates~\cite{Theis+:16,Levine+:18}. As a result, neutral-atom-based platforms currently allow for controlled quantum 
dynamics of more than $100$ qubits, with the prospect of reliable QIP with even much larger qubit systems~\cite{Saffman:16,Henriet+:20,ShiREVIEW:22} deemed to be realistic in the not-too-distant future~\cite{Morgado+Whitlock:21}. This has, in turn, reinvigorated 
research interest in quantum-state engineering in this class of atomic systems~\cite{Buchmann+:17,Ostmann+:17,Malinovskaya:17, 
Omran+:19,Zheng+:20,Wintermantel+:20,Mukherjee+:20,Haase+:21,Pachniak+Malinovskaya:21}.

An ordinary (prototype) $N$-qubit $W$ state~\cite{Koashi+:00} is an equal superposition of all $N$-qubit basis states with exactly 
one qubit in their ``up'' state, and all the remaining ones in their ``down'' state. In systems with a periodic spatial arrangement 
(i.e. a lattice)~\cite{Hauck+Stojanovic:22} of qubits it makes sense to consider generalized $W$ states, which represent linear combinations of the same $N$-qubit basis states but with a constant phase difference between contributions corresponding to adjacent lattice sites. 
This phase difference corresponds to a quasimomentum from the Brillouin zone of the underlying lattice. The physical meaning of 
such states -- which in the following will be refereed to as {\em twisted} $W$ states -- becomes fully transparent upon switching 
from spin-$1/2$ to spinless-fermion degrees of freedom using the Jordan-Wigner transformation~\cite{ColemanBOOK:15}. Namely, 
these states are equivalent to Bloch states of spinless-fermion excitations.

Aside from realizing $W$~\cite{StojanovicPRA:21} and GHZ states~\cite{Omran+:19}, interconversion between those states 
is another relevant problem of quantum-state engineering. This problem was first addressed in the context of a photonic 
system~\cite{Walther+:05}, where such interconversion can be carried out only in a nondeterministic fashion. In recent 
years the same problem was investigated in a system of three equidistant Rydberg atoms with van-der-Waals-type interaction,
which are at the same time acted upon by several external laser fields~\cite{Zheng+:20,Haase+:21}. This last system was first
studied~\cite{Zheng+:20} using the method of shortcuts to adiabaticity~\cite{STA_RMP:19}, more precisely Lewis-Riesenfeld 
invariants~\cite{Lewis+Riesenfeld:69}, followed by an alternative treatment~\cite{Haase+:21} that made use of a dynamical-symmetry-based 
approach~\cite{BarutBOOK:71}.

In this paper, we consider the system of three neutral atoms in the Rydberg-blockade (RB) regime~\cite{Tong+:04,Singer+:04,Urban+:09,Gaetan+:09}, 
interacting with external laser fields, with our research objective being twofold. We first present a deterministic preparation 
scheme for {\em chiral} $W$ states, a special class of twisted three-qubit $W$ states that are of relevance for implementing 
noiseless-subsystem qubit encoding~\cite{Knill+:00,Cole+:21}. We then address the problem of deterministically converting twisted 
$W$ states into their GHZ counterparts via different intermediate states. Both of these dynamical generation schemes rely heavily 
on relative alignments of the laser fields involved and precise positioning of the atoms, these ingredients being within reach of 
nowadays' technology~\cite{Yu+:18,Norcia+:18,Liu+:19,Deist+:22}. Furthermore, we show that even without such an experimental precision 
a conversion of a twisted $W$ state is still possible. A strong laser driving field, which introduces light shifts, can determine the 
specific twisted states participating in a conversion scheme. The scheme also makes use of additional weaker fields, which address 
the lifted degeneracies of internal energy levels of the Rydberg trimer.

Our principal result in the context of the state-conversion problem -- relative to previous studies of this problem~\cite{Zheng+:20,Haase+:21} 
that concentrated only on ordinary $W$ states -- is that starting from twisted $W$ states is equivalent to renormalizing downwards the 
relevant Rabi frequencies of external laser pulses. This renormalization, which depends only on the relative alignment between 
the laser fields used, leads to somewhat longer state-conversion times (for the same laser-pulse energy used) than in the case with 
ordinary $W$ states. However, we also demonstrate that the latter times are at least two orders of magnitude shorter than the typical 
lifetimes of relevant Rydberg states.

The remainder of this paper is organized as follows. In Sec.~\ref{sec:system} we introduce the neutral-atom system under consideration 
and briefly describe its interaction with external laser fields. In Sec.~\ref{sec:generalized_states} we introduce several classes of entangled 
multiqubit states of interest for the present work ($W$-, GHZ-, and Dicke states) and the notation to be used throughout the paper. 
Section~\ref{sec:effH} is devoted to the derivation of effective Hamiltonians of the system that serve as the point of departure 
for the state-engineering schemes discussed in the present work. In Sec.~\ref{sec:preptwistedW} we provide a discussion of specific alignments 
of laser fields and relative atom positions that are required for the generation of twisted $W$ states from the atomic-ensemble ground state 
via $\pi$-pulses of a single laser field resonant with the Rydberg transition. In particular, we describe in detail a preparation scheme for 
chiral $W$ states in a Rydberg trimer. In Sec.~\ref{sec:StateEngineer} we present two different schemes for the conversion of twisted 
$W$ states into GHZ states, which are respectively based on degenerate Dicke manifolds of states and lifted degeneracies. 
We conclude, with a summary of the obtained results and a short survey of possible directions for future investigation, 
in Sec.~\ref{sec:conclusion}. For the sake of completeness, some relevant mathematical details are presented in detail
in Appendices~\ref{app:effH} and \ref{app:twistcompensation}.
	
\section{System and atom-field interaction} \label{sec:system}
We consider a system that consists of $N$ identical neutral atoms (e.g. of $^{87}$\:Rb) located at positions determined by 
the vectors $\vec{x}_n$ ($n=1,2,\ldots,N$).
Anticipating the use of external laser pulses that are all close to being resonant with the same internal atomic transition -- namely, 
the one between the ground state $\ket{g}_n$ (with energy $E_g$) and a highly-excited Rydberg state $\ket{r}_n$ (energy $E_r$) -- the 
atoms can be treated as effective two-level systems with the atomic frequency $\omega_A=(E_r-E_g)/\hbar$ as resonance frequency.
In the following we will treat $E_g$ as the origin of the energy scale, i.e. set $E_g=0$.
In the QIP context, each atom in this system represents a $gr$-type qubit~\cite{Morgado+Whitlock:21}, where the atomic states $\ket{g}_n$ 
and $\ket{r}_n$ play the role of the logical ``down'' ($\ket{0}_n$) and ``up'' ($\ket{1}_n$) states of the $n$-th qubit, respectively.
Recalling that the typical energy splitting of $gr$-type qubits is in the range between $900$ and $1500$ THz~\cite{Morgado+Whitlock:21}, 
manipulations of such qubits require either an ultraviolet laser or a combination of visible and infrared lasers in a ladder configuration.

\begin{figure}[t!]
\includegraphics[width=\linewidth]{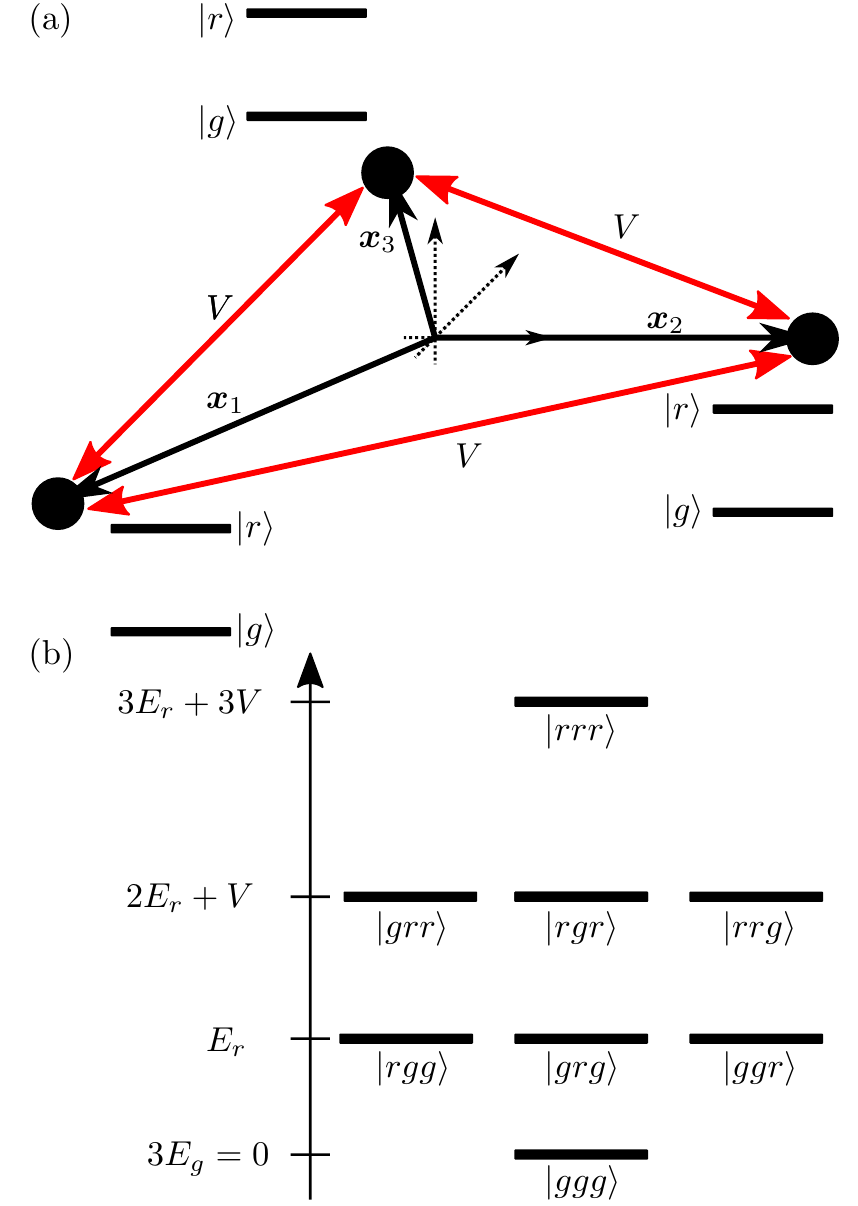}
\caption{\label{fig:systemhamiltonian}
(a) Schematic illustration of a Rydberg-atom trimer ($N=3$).
The three Rydberg atoms, located at the positions specified by the vectors $\vec{x}_n$ ($n=1,2,3$), 
form an equilateral triangle. The ground- and Rydberg states of each atom are denoted by $\ket{g}$ 
and $\ket{r}$, respectively, while $V$ stands for the magnitude of their pairwise van der Waals 
interaction.\\
(b) Energy-level scheme of a Rydberg-atom trimer. The origin of the energy scale is chosen such that $E_g=0$. 
We are considering long interaction times $T$, i.e. $|E_r -E_g| T/\hbar \gg1$, $|V| T
/\hbar \gg1$, and large electronic excitation energies, i.e. $| E_r -E_g| \gg|V|$.}
\end{figure}

We also assume that the atoms are pairwise coupled through off-resonant dipole-dipole (van der Waals) interaction. In the special case 
of equidistant atoms -- the physical situation of primary relevance in the remainder of the present work -- the 
magnitude $V_{pq}=C_6/d_{pq}^6$ of this interaction (where $d_{pq}$ is the distance 
between atoms $p$ and $q$ and $C_6$ the van der Waals interaction constant) is the same for all pairs 
of atoms, and we denote $V_{pq}=V$. For $N=3$ the case of equidistant atoms corresponds to their arrangement 
in the form of an equilateral triangle [for a pictorial illustration, see Fig.~\ref{fig:systemhamiltonian}(a)], while for $N=4$
they are located at the vertices of a regular tetrahedron.

Importantly, our envisioned system is also assumed to be in the RB regime~\cite{Tong+:04,Singer+:04,Urban+:09,Gaetan+:09}, 
which is equivalent to demanding that the interaction-induced energy shift $V$ far exceeds the Fourier-limited width 
of all the utilized laser pulses (i.e. $|V| T_{\textrm{int}}/\hbar \gg 1$, where $T_{\textrm{int}}$ is the pulse duration).
Thus, the state-preparation- and conversion schemes to be presented in what follows are applicable in the regime of primary interest for 
QIP, as the phenomenon of RB provides the conditional logic that enables neutral-atom quantum computing~\cite{Shi:18}.
The suitability of our envisioned system for quantum-state engineering is further underscored by its reliance on $gr$-type qubits, 
which -- owing to their straightforward initialization, manipulation, and measurements -- represent the preferred neutral-atom qubit 
type for fast, high-fidelity entangling operations~\cite{Morgado+Whitlock:21}.

The total Hamiltonian of the system at hand is given by $H=H_A+H_F+H_{\textrm{int}}$, where $H_A$ describes the atomic ensemble, 
$H_F$ the free external fields, and $H_{\textrm{int}}$ the atom-field interaction.
The form of these three contributions to the total system Hamiltonian will be discussed in detail in the following.

The Hamiltonian of the atomic ensemble is given by
\begin{align} \label{eq:H_A}
H_A =& \sum_{n=1}^N \hbar\omega_A \ket{r}_{nn}\bra{r}
+ \sum_{p<q}^N V \ket{r}_p\ket{r}_{qq}\bra{r}_p\bra{r} \:.
\end{align}
The energy eigenvalues of the atomic ensemble are given by $E_a = a\,\hbar\omega_A + V \uber{a}{2}$, 
where $a\leq N$ is the number of atoms in the excited state.
The energy level $E_a$ has a degeneracy of $\binom{N}{a}$ and the energy gap between adjacent excitation subspaces is given 
by $\Delta E_a \equiv E_{a}-E_{a-1} = \hbar\omega_A + V(a-1)$.
In particular, the energy-level scheme in the Rydberg-trimer case ($N=3$) is pictorially illustrated in Fig.~\ref{fig:systemhamiltonian}(b).

The atomic ensemble is subject to multiple plane-wave laser fields with respective wave vectors 
$\vec{k}_j$ and frequencies $\omega_j$ ($j=0,1,\ldots,J$). The fields are quantized using creation 
and annihilation operators $a_j^\dagger$ and $a_j$, such that the free-field Hamiltonian is given by
\begin{equation}
H_F = \sum_{j=0}^J\hbar\omega_j a^\dagger_j a_j \:,
\end{equation}
where -- for the sake of convenience -- the ground-state energies $\hbar\omega_j/2$ of all modes are omitted.
We treat the interaction between laser pulses and the atomic ensemble in the dipole- and rotating-wave approximations (RWA), 
with the corresponding coupling strengths being denoted by $d_j$. All laser fields are assumed to resemble classical fields. 
Hence, they can be described as coherent field states~\cite{Glauber:63} of high mean photon numbers $M_j$, such 
that the coupling constants define (resonant) Rabi frequencies $\Omega_j = \sqrt{M_j}\:d_j/\hbar$. 
Here $d_j = -i \sqrt{\hbar\omega_j/(2\epsilon_0\tilde{V})}\:\langle r|{{\bf d}}\cdot \bf{\epsilon}_j|g\rangle$, 
where $\epsilon_0$ is the vacuum dielectric constant, $\tilde{V}$ is the quantization volume of the 
field modes, $\bf{\epsilon}_j$ is the polarization unit vector of mode $j$ (orthogonal to its propagation 
direction), and ${{\bf d}}$ is the atomic dipole operator.

The atom-field interaction in this system is described by the Hamiltonian
\begin{equation} \label{eq:H_int1}
H_\textrm{int}=\sum_{n=1}^N \sum_{j=1}^J \left(\ket{g}_{nn}\bra{r} 
d^*_j a^\dagger_j{e^{-i \vec{k}_j\cdot\vec{x}_n}} +\textrm{H.c.}\right) \:,
\end{equation} 
where the site-dependent phases $\vec{k}_j\cdot\vec{x}_n$ result from evaluating the mode function for plane waves 
at the distinct atom positions. This Hamiltonian can be recast in the form
\begin{equation} \label{eq:H_int}
H_\textrm{int}=\sum_{n=1}^N \sum_{j=1}^J\left[ U(\vec{k}_j)\ket{g}_{nn}\bra{r} 	
d^*_j a^\dagger_j U^\dagger(\vec{k}_j) +\textrm{H.c.} \right]
\end{equation} 
by introducing the transformation
\begin{equation}\label{defUk}
U(\vec{k}) = \bigotimes_{n=1}^N
\left(e^{i\vec{k}\cdot\vec{x}_n}\ket{r}_{nn}\bra{r}+\ket{g}_{nn}\bra{g}\right) \:.
\end{equation}
In the following the transformation $U(\vec{k})$ will be used repeatedly in order to simplify the description of 
the system under consideration. For the sake of brevity, we will just use $\vec{k}$ to parametrize this transformation, 
because the atoms are assumed to be located at fixed positions.

\section{Relevant multiqubit states} \label{sec:generalized_states}
In what follows, we introduce several classes of entangled multiqubit states of interest in the remainder of this 
work. In Sec.~\ref{WGHZDicke} we discuss generalized $W$ states, together with their GHZ counterparts. We also briefly 
introduce Dicke states and their twisted counterparts that play an auxiliary role in our further discussion. 
In Sec.~\ref{ChiralWintro} we specialize to the three-qubit systems, introducing first the chiral $W$ states, and 
then explaining their connection to specific twisted $W$ states in the system under consideration (Rydberg trimer).
The notation used will be the one appropriate for $gr$-type Rydberg-atom qubits \cite{Shi:19}, with $\{|g\rangle,|r\rangle\}$ being
the relevant computational basis of a single qubit.
\subsection{Generalized $N$-qubit $W$-, GHZ-, and Dicke states} \label{WGHZDicke}
The most general $W$-type states, not necessarily maximally entangled, represent linear combinations of states 
in which exactly one qubit is in the state $|r\rangle$, with all the remaining ones being in the state $|g\rangle$ 
(i.e. all states corresponding to Hamming-weight-$1$ bit strings). They are given by
\begin{equation}\label{defGeneralW}
|W_N (A_1,\ldots, A_N)\rangle = \frac{1}{\sqrt{A}}\sum_{n=1}^{N}A_n|g\ldots r_n \ldots g\rangle \:,
\end{equation}
where $A=\sum_{n=1}^{N}|A_n|^2$ and $A_n$ are $N$ arbitrary complex numbers, with at least two of them being unequal 
to zero. In the special case with $|A_n|=1/\sqrt{N}$ ($n=1,\ldots,N$), one can substitute $A_n=e^{i\phi_n}/\sqrt{N}$ 
and recast the last states in the form
\begin{equation}\label{defPhaseW}
|W_N (\phi_1,\ldots, \phi_N)\rangle = \frac{1}{\sqrt{N}}
\sum_{n=1}^{N}e^{i\phi_n}|g\ldots r_n \ldots g\rangle \:.
\end{equation}
Here the phases $\phi_1,\ldots,\phi_N$ are defined modulo $2\pi$, and -- as usual -- the state $|W_N\rangle$ is defined up to 
a global phase. In particular, the special case $\phi_1=\ldots=\phi_N=0$ of the latter maximally-entangled states are the 
most often used, ``ordinary'' $W$ states
\begin{equation}\label{eq:defOrdinaryW}
|W_N\rangle = \frac{1}{\sqrt{N}}\sum_{n=1}^{N}\:|g\ldots r_n \ldots g\rangle \:.
\end{equation}

The most notable property of $W$ states is that they are the most robust ones to particle loss 
among all $N$-qubit states~\cite{Koashi+:00}.
The entanglement inherent to $W$ states is fundamentally different than that of GHZ states
\begin{equation}
\ket{\text{GHZ}_N} = \frac{1}{\sqrt{2}}(\ket{rr\ldots rr}+e^{i\phi}\ket{gg\ldots gg}) \:,
\end{equation}
no matter whether one considers pairwise- or distributed entanglement. For instance, in the $N=3$ case
the $W$ state are characterized by a strong pairwise entanglement (as quantified by the corresponding concurrences)
while the essential three-way entanglement (as quantified by the $3$-tangle) vanishes~\cite{Coffman+:00}.
On the other hand, its GHZ counterpart has maximal essential three-way entanglement, while pairwise entanglements 
vanish~\cite{HorodeckiRMP:09}.

In the case of a periodic spatial arrangement (lattice) of qubits, it is pertinent to introduce the ``twisted'' $W$ 
states as a special case of the states in Eq.~\eqref{defPhaseW} where $\phi_n\equiv\vec{k}\cdot\vec{x}_n$, with 
$\vec{k}$ being a quasimomentum from the Brillouin zone corresponding to the underlying lattice of qubits with positions 
$\vec{x}_n$. Those states are given by
\begin{equation}\label{defTwistW}
|W_N(\vec{k})\rangle = \frac{1}{\sqrt{N}}\sum_{n=1}^{N}e^{i\vec{k}\cdot\vec{x}_n}
|g\ldots r_n \ldots g\rangle \:.
\end{equation}
For instance, if qubits form a regular one-dimensional lattice, then the quasimomentum, expressed in units of the inverse 
lattice period, belongs to $(-\pi,\pi]$.
The special significance of the state in Eq.~\eqref{defTwistW} rests on the notion that using Jordan-Wigner transformation 
from pseudospin-$1/2$ to spinless-fermion (or hardcore-boson) degrees of freedom~\cite{ColemanBOOK:15} this state is mapped 
onto a bare-excitation Bloch state with quasimomentum $\vec{k}$.
In particular, the ordinary $W$ state with $\phi_1=\phi_2=\ldots=\phi_N=0$ -- the special case of Eq.~\eqref{defTwistW} with 
$\vec{k}\cdot\vec{x}_n=0$ -- corresponds to the $\vec{k}=0$ Bloch state.

To describe all possible $N$-qubit states we consider different numbers $a$ of excited qubits.
A generic state in the subspace of states with $a$ excitations can be parameterized as $\ket{\{n_1,\ldots,n_a\}}$, where the 
atoms enumerated with $n_1,\ldots,n_a$ are in the excited state while the remaining ones are in the ground state.
The Dicke state
\begin{equation}
\ket{D^{N}_{a}}= \uber{N}{a}^{-1/2}\sum_{n_1<\ldots<n_a}^N\ket{\{n_1,\ldots,n_a\}} \:
\end{equation}
represents the equal superposition of all states $\ket{\{n_1,\ldots,n_a\}}$ spanning that subspace, 
where the sum in the last equation runs over all $\uber{N}{a}$ combinations of $a$ atoms out of $N$.

In a completely analogous way as in the case of $W$ states, one can introduce twisting, i.e. $\vec{k}$-dependent 
relative phases between different $N$-qubit basis states with equal excitation number $a$.
These phases are captured by the unitary transformation of Eq.~\eqref{defUk}, which maps the Dicke states into 
their twisted counterparts
\begin{equation}
\ket{D^N_a(\vec{k})}=
\sqrt{\frac{1}{\uber{N}{a}}}\sum_{n_1<\ldots<n_a}^N
e^{i\vec{k} \cdot \sum_{l=1}^a \vec{x}_{n_l}}\ket{\{n_1,\ldots,n_a\}} \:.
\end{equation}
Obviously, twisted $W$ states are a special case ($a=1$) of twisted Dicke states, i.e. 
$\ket{W_N(\vec{k})}\equiv\ket{D_1^N(\vec{k})}$.
\subsection{Chiral three-qubit $W$ states} \label{ChiralWintro}
Given that all of our numerical calculations in the following will pertain to the $N=3$ case, it is prudent to devote 
special attention to three-qubit systems and introduce a special notation
that allows one to conveniently denote the three-qubit states. For this purpose we introduce
the self-adjoint {\em chirality} operator~\cite{Viola+:01}
\begin{equation} \label{chirality}
\chi = \frac{1}{2\sqrt{3}}\sum_{\alpha,\beta,\gamma}
\epsilon_{\alpha\beta\gamma}\:\sigma_{1\alpha}
\sigma_{2\beta}\sigma_{3\gamma} \:.
\end{equation}
Here the indices $\alpha$, $\beta$, and $\gamma$ run over $x$, $y$, and $z$, with $\epsilon_{\alpha\beta\gamma}$ 
being the totally antisymmetric Levi-Civita symbol defined in terms of these indices.
$\sigma_{n\alpha}$ is a Pauli operator acting on qubit $n$ ($n=1,2,3$). Orthonormal eigenstates $|\zeta_{as}\rangle$  
of this chirality operator are explicitly given by
\begin{equation} \label{Wlike}  \begin{aligned}
|\zeta_{00}\rangle &= |ggg\rangle  \:,  \\
|\zeta_{10}\rangle &= (|rgg\rangle + |grg\rangle +|ggr\rangle)/\sqrt{3}\:,  \\
|\zeta_{1+}\rangle &= (w^{*}|rgg\rangle + |grg\rangle +w|ggr\rangle)/\sqrt{3}\:,  \\
|\zeta_{1-}\rangle &= (w|rgg\rangle + |grg\rangle +w^{*}|ggr\rangle)/\sqrt{3}\:, \\
|\zeta_{20}\rangle &= (|grr\rangle + |rgr\rangle +|rrg\rangle)/\sqrt{3}\:, \\
|\zeta_{2+}\rangle &= (w^{*}|grr\rangle + |rgr\rangle +w|rrg\rangle)/\sqrt{3}\:, \\
|\zeta_{2-}\rangle &= (w|grr\rangle + |rgr\rangle +w^{*}|rrg\rangle)/\sqrt{3}\:, \\
|\zeta_{30}\rangle &= |rrr\rangle  \:, 
\end{aligned} \end{equation}
with $w \equiv\exp(2\pi i/3)$~\cite{Cole+:21}. They constitute a basis of the state space of three qubits. 
The quantum number $a=0,\ldots,3$ denotes the number of qubits in the $|r\rangle$ state (i.e. the Hamming weight of 
the corresponding bit string), and the additional quantum number $s = 0, \pm$ identifies the eigenstates uniquely.
Among these eight basis states there are three $W$ states, i.e. states corresponding to Hamming-weight-$1$ 
bit strings ($a = 1$) -- the ordinary $W$ state $|\zeta_{10}\rangle \equiv |W_{3}(\vec{k}=0)\rangle$ and two 
{\em chiral} $W$ states $|\zeta_{1+}\rangle$ and $|\zeta_{1-}\rangle$.
The mutual orthogonality of these three $W$ states is thus a consequence of the fact that they belong to three 
different eigensubspaces of the chirality operator [cf. Eq.~\eqref{chirality}].

It is important at this point to establish a connection between the two chiral $W$ states and the general 
twisted three-qubit states [as defined by Eq.~\eqref{defTwistW}], which is of interest for our treatment 
of the Rydberg-atom system under consideration. Our assumed spatial arrangement of three neutral-atom qubits 
in the form of an equilateral triangle [cf. Fig.~\ref{fig:systemhamiltonian}(a)], which implies that these three 
qubits are symmetrically positioned on a circle, is equivalent to an array (i.e. a one-dimensional lattice) 
of three qubits with periodic boundary 
conditions imposed. In particular, it is straightforward to verify that the state $|\zeta_{1+}\rangle$ is -- up 
to an irrelevant global phase -- equivalent to the twisted state $|W_3(k=2\pi/3)\rangle$ of a three-qubit array 
that corresponds to the quasimomentum $k=2\pi/3$ (expressed in units of the inverse lattice spacing). Similarly, 
$|\zeta_{1-}\rangle$ is equivalent to the twisted state $|W_3(k=-2\pi/3)\rangle$. Having established the correspondence 
between the two chiral states and the twisted $W$ states of a one-dimensional array of three qubits, we will in 
the following use $|\zeta_{1+}\rangle$ and $|\zeta_{1-}\rangle$ as our primary examples for the latter class of 
generalized three-qubit $W$ states.

\section{Effective Hamiltonians} \label{sec:effH}
In what follows, we describe the derivation of effective system Hamiltonians that constitute the basis for 
designing various state-engineering schemes in the present work. These effective Hamiltonians, which are 
derived with reference to twisted (rather than ordinary) $W$ states, constitute a generalization 
of the effective four-level Hamiltonian that was first presented in Ref.~\cite{Zheng+:20}. In particular,
we first treat the case of resonant laser fields (Sec.~\ref{ssec:offresonant}), followed by a discussion 
of the off-resonant ones (Sec.~\ref{ssec:resonant}).

To realize different state-preparation- and conversion schemes in the neutral-atom system under consideration, we 
derive different effective Hamiltonians using the resolvent formalism (see, e.g., Ref.~\cite{Galindo+PascualBOOK:91}).
The effective Hamiltonian can most generally be written in the form
\begin{equation}\label{eq:1stHeff}
H_{\textrm{eff}} = \sum_{E\in \sigma(H_0)}P_E\left(H + HQ_E\frac{1} {E-Q_E H Q_E} Q_E H\right)P_E\:,
\end{equation}
where the sum runs over the whole energy spectrum $\sigma(H_0)$ of the non-interacting part $H_0=H_A+H_F$ of the 
total Hamiltonian of the system.
This equation is a direct consequence of the general relation for the resolvent $G(E)$ of the Hamiltonian $H$, i.e. 
$G(E) = (E-H)^{-1}$ and of the definition of the effective Hamiltonian $H_{\textrm{eff}}$ in terms of the orthogonal 
projection operators $P_E$ and $Q_E = \mathds{1} - P_E$, i.e. $P_E G(E)P_E = (E-H_{\textrm{eff}})^{-1}P_E$. Thus, the 
effective Hamiltonian $H_{\textrm{eff}}$ describes the dynamics  inside the subspace of the Hilbert space only which 
is characterized by the projection operator $P_E$. By choosing the orthogonal projection operators $P_E$ and $Q_E$ 
appropriately, effective Hamiltonians can be determined systematically within a perturbative framework so that secular 
terms are avoided in the time evolution. If a single mode is considered, the (unperturbed) energy eigenvalues are 
$E_a^m=E_a+m\hbar\omega$, where $a$ is the number of excited atoms and $m$ is the photon number of this mode, and the 
projection operators of  Eq.~\eqref{eq:1stHeff} project onto the corresponding (degenerate) energy subspaces. For a 
perturbative approach to first order, the denominator of the resolvent of  Eq.~\eqref{eq:1stHeff} can be approximated 
by the unperturbed Hamiltonian.

The concrete form of the projectors $P_E$ depends on the considered fields. We discuss two distinct cases -- off-resonant 
and resonant laser fields. The crucial difference between them is that off-resonant laser fields do not introduce additional 
energy degeneracies, i.e. $E_a^m=E_{a'}^{m'}$ if and only if the two atomic-excitation numbers are the same ($a=a'$) and the 
two photon numbers as well ($m=m'$). As long as degeneracies due to different fields are well separated, we can split up the 
sum over all fields in the system Hamiltonian $H$ and treat each field separately. By tracing out the field's degrees of 
freedom we will derive effective Hamiltonians describing the dynamics of the atomic ensemble via corrections to the atomic 
ensemble Hamiltonian $H_A$. In the following we will discuss the two cases of a single laser field separately.
The resulting effective Hamiltonians and their combinations will then be used in Secs.~\ref{sec:preptwistedW} 
and~\ref{sec:StateEngineer} for state preparation and conversion schemes, respectively.

\subsection{Off-resonant laser field} \label{ssec:offresonant}
We first consider a single off-resonant laser field (enumerated with $j=0$) with wave vector $\vec{k}_0$, assuming that its 
detuning $\Delta_{0}=\omega_0-\omega_A \equiv \omega_0-(E_r-E_g)/\hbar$ is much larger in absolute value than the corresponding 
Rabi frequency $\Omega_0$, i.e. $|\Delta_{0}|\gg |\Omega_0|$.

Because the field is assumed to be off-resonant, the degenerate energy subspaces are completely determined by the number of 
excitations in the atomic ensemble $a$ and the number of laser-field excitations, i.e. the number of photons $m_0$. The projectors 
onto a subspace of energy $E_a^{m_0}$ is given by
\begin{align} \label{eq:Poff}
P^{m_0}_a= \sum_{n_1<\ldots<n_a}^N & \ket{\{n_1,\ldots ,n_a\}} \, 
\bra{\{n_1,\ldots ,n_a\}}
 \nonumber \\ &
 \otimes \ket{m_0}\bra{m_0} \:.
\end{align}
A detailed derivation of the effective Hamiltonian [cf. Eq.~\eqref{eq:1stHeff}] is relegated to Appendix~\ref{app:effHoffresonant}. 
Here we only state the resulting corrections to the Hamiltonian $H_A$ of the atomic ensemble, which are obtained by assuming 
a coherent field state of high mean photon number $M_0$ and tracing out the field degrees of freedom. To succinctly write down 
these corrections, we make use of the operator
\begin{equation}
\operatorname{Hd}_2(N) = \sum_{n_1,n_2}^N (1-\delta_{n_1n_2})\ket{r}_{n_1n_1}
\bra{g}\otimes\ket{g}_{n_2n_2}\bra{r} \:,
\end{equation}
which transforms every state into an equal (not necessarily normalized) superposition of all other states connected to it via 
precisely one excitation and one de-excitation at different atoms. In other words, speaking in terms of bit strings with $g\equiv 0$ 
and $r\equiv 1$, this operator connects all bit strings of equal Hamming weight but with a Hamming distance (Hd) of two.

The lowest-order corrections to the atomic ensemble Hamiltonian are captured by 
\begin{align} \label{eq:effHoff}
H^{\text{off}}_N(\vec{k}_0) = \sum_{a=0}^{N} &\hbar^2|\Omega_0|^2 P_a \nonumber \\
&\times \left[\frac{ U(\vec{k}_0)
\operatorname{Hd}_2(N)U^\dagger(\vec{k}_0)+N-a}{\hbar\Delta_0- a V}\right. \nonumber\\
 & \phantom{\times\Big[}
\left.-\frac{U(\vec{k}_0)
\operatorname{Hd}_2(N) U^\dagger(\vec{k}_0)+a}{\hbar\Delta_0- (a-1) V}\right]P_a \:,
\end{align}
such that the effective Hamiltonian of the atomic ensemble is given by $H_{\textrm{eff}}=
H_A+H^{\text{off}}_N(\vec{k}_0)$.
In the last equation
\begin{equation}
P_a = \sum_{n_1<\ldots<n_a}^N \ket{\{n_1,\ldots ,n_a\}} \, \bra{\{n_1,\ldots ,n_a\}}
\end{equation}
stands for the projector onto the subspace of $a$ excitations, while the effect of the site-dependent phase shifts
is captured by the unitary-transformation operators $U(\vec{k}_0)$ [cf. Eq.~\eqref{defUk}].

It is important to note that for $a=0,1,N-1,N$ the corresponding terms in the Hamiltonian of Eq.~\eqref{eq:effHoff} can 
further be simplified using the following identities:
\begin{align}
P_0\operatorname{Hd}_2(N)P_0 &= P_N\operatorname{Hd}_2(N) P_N = 0 \:, \nonumber \\
P_1\operatorname{Hd}_2(N)P_1&=N\ket{D^N_1}\bra{D^N_1}-P_1 \:,\\ 
P_{N-1}\operatorname{Hd}_2(N)P_{N-1}&=N\ket{D^N_{N-1}}\bra{D^N_{N-1}}-P_{N-1} \nonumber \: .
\end{align}

For Rydberg trimers ($N=3$) all terms of $\operatorname{Hd}_2$ reduce to one of the above special cases and and we obtain
\begin{align} \label{eq:effH_offres3}
H^{\text{off}}_{3}(\vec{k}_0) 
= {} 			&  3 s_0 \ket{ggg}\bra{ggg} - 3 s_2\ket{rrr}\bra{rrr} \nonumber
\\ \nonumber 	& +(-3s_0+3s_1)\ket{D^3_1(\vec{k}_0)}\bra{D^3_1(\vec{k}_0)}
\\ \nonumber 	& +(-3s_1+3s_2){\ket{D^3_2(\vec{k}_0)}	\bra{D^3_2(\vec{k}_0)}}
\\ 				& + s_1 \left(P_1-P_2\right) \: ,
\end{align}
where $s_a=\hbar^2|\Omega_0|^2/(\hbar\Delta_0-aV)$ is a shorthand for the energy shifts.
Hence, the off-resonant laser field just shifts the energy levels of the atomic ensemble, but up to first order does not 
contribute any off-diagonal elements.
This result does not include jump operators between different levels.
They would appear in higher order terms of the resolvent expansion, but are not considered here.
Therefore, we neglect small oscillatory behavior in the level populations of the atomic ensemble of the order 
of $\max\{[s_a/(\hbar|\Omega_0|)]^2; 0 \leq a \leq N-1\}$.
Due to the induced energy shifts, the effective Hamiltonian lifts the energy degeneracies of the subspaces with 
$a=1,2$ excitations, such that $\ket{D^3_1(\vec{k})}$ and $\ket{D^3_2(\vec{k})}$ differ in energy from the corresponding 
orthogonal states of the same total excitation number $a$.
A suitable eigenbasis of this effective Hamiltonian are the $\ket{\zeta_{as}}$ states in Eqs.~\eqref{Wlike}. 
The energy shifts can be set to drive specific transitions by choosing appropriate detunings of additional fields as 
will be discussed in Sec.~\ref{ssec:twistcompensation} below.

\subsection{Resonant laser fields} \label{ssec:resonant}
If a field (enumerated $j=a$) is in resonance with a specific transition $a\leftrightarrow a-1$ of the atomic ensemble, 
the subspaces $P_a^{m_a}$ and $P_{a-1}^{m_a+1}$ become energetically degenerate.
By equating the energies $E_a^{m_a}$ and $E_{a-1}^{m_a+1}$, one obtains the condition 
\begin{equation}
\hbar\Delta_a = \hbar(\omega_a-\omega_A) = (a-1) V\:,
\end{equation} 
where $\omega_a$ is the frequency of the laser field. Given this degeneracy, the last two subspaces 
have to be jointly considered within the framework of the resolvent formalism.
Assuming the field to be classical, i.e. in a coherent state of high mean photon number $M_a$, and tracing over the degrees 
of freedom of the field, we obtain an effective Hamiltonian for the atomic system alone.
If we further assume that $V\gg \hbar |\Omega_a|$, we can neglect all terms scaling with $|\hbar\Omega_a|^2/V$, thus 
leaving -- apart from $H_A$ -- just parts containing the subspace ladder operator 
\begin{equation}
\sigma_a^- = \sum_{n_1<\ldots <n_a}^N \sum_{n=1}^N \ket{g}_{nn}\braket{r}
{\{n_1,\ldots ,n_a\}}\bra{\{n_1,\ldots ,n_a\}}  
\end{equation} 
and its Hermitian conjugate.
This results in the effective Hamiltonian $H_A+H^{a\leftrightarrow a-1}_N$, where
\begin{align} \label{eq:effHres}
H^{a\leftrightarrow a-1}_N(\vec{k}_a) &= \hbar\Omega_a^* \: U(\vec{k}_a)
\sigma^-_aU^\dagger(\vec{k}_a) + \textrm{H.c.}\:.
\end{align}
In the last equation, whose detailed derivation is presented in Appendix~\ref{app:effHresonant}, the effect of 
the site-dependent phase shifts is once again encoded into the unitary transformation $U(\vec{k}_a)$.
The special case $a=1$ corresponds to the well-known effect of enhanced Rabi oscillations~\cite{Morgado+Whitlock:21} 
and leads to a simple preparation scheme for twisted $W$ states, as discussed in Sec.~\ref{sec:preptwistedW} below.

Combining $N$ laser fields with detunings $\Delta_a=(a-1)V$ with $a\in\{1,\ldots ,N\}$, such that every laser field 
is in resonance with one specific transition between eigenstates of the atomic Hamiltonian $H_A$, we can construct 
an effective Hamiltonian connecting stepwise all $N+1$ degenerate energy levels of the atomic ensemble.
The corrections added to $H_A$ in this case are
\begin{equation} \label{eq:effHLadder}
H^\text{L}_N(\{\vec{k}_a\}) = \sum_{a=1}^N [\hbar\Omega_a^*
U(\vec{k}_j)\sigma_a^- U^\dagger(\vec{k}_a) + \textrm{H.c.}] \:,
\end{equation}
where $\{\vec{k}_a\}=\{\vec{k}_1,\dots,\vec{k}_N\}$ and each laser field $j=1,\ldots,N$ connects the subspace 
of $a'=j-1$ and $a=j$ excitations like a step on a ladder (L).
These fields in general have different corresponding wave vectors $\vec{k}_a$, thus introducing different site-dependent 
phase shifts. Because in general $U(\vec{k}_a)\neq U(\vec{k}_{a\pm 1})$ these fields do not necessarily form a ladder Hamiltonian 
of $N+1$ states. To what extent the steps match is described by the overlaps
\begin{multline}
\braket{D^N_a(\vec{k}_a)}{D^N_a(\vec{k}_{a+1})} \\ 
=\uber{N}{a}^{-1} \sum_{n_1<\ldots <n_a}^N e^{i(\vec{k}_{a+1}-\vec{k}_a)\cdot\sum_{j=1}^a \vec{x}_{n_j}} \: .
\end{multline}
Overlaps smaller than unity result in offsets. One way to deal with the latter is to control laser alignments and set 
precise atom positions, such that special atomic ensemble states with their specific relative phase are selected in 
the effective Hamiltonian. The easiest case is to avoid phase differences between different atom positions in the first place.
If all laser fields are properly aligned such that $\vec{k}_a\cdot \vec{x}_n=0 \mod 2\pi$ for all combinations of $a,n=1,\dots,N$, 
the effective Hamiltonian is characterized by perfect overlaps and connects all $N+1$ different $\ket{D^N_a}$ states 
succeedingly.
The possibility to select different states for the state conversion is discussed in Sec.~\ref{ssec:simplifiedWtoGHZ}.
Alternatively, such a strong off-resonant field can be used to lift some of the degeneracies in $H_A$.
The energy shifts introduced by this field ($j=0$) can be used to select certain parts of the Hamiltonian dynamics by a 
fine detuning of the ladder fields $j=1,\ldots ,J$. Such compensation of unwanted terms will be carried out for the Rydberg-trimer 
case in Sec.~\ref{ssec:twistcompensation}.

\section{Preparation of chiral $W$ states} \label{sec:preptwistedW}
In this section we present a state-preparation scheme for twisted $W$ states for the Rydberg system under consideration by 
making use of the effective Hamiltonian for a single resonant laser field derived in Sec.~\ref{ssec:resonant}.
It includes the preparation of the chiral states $\ket{\zeta_{1+}}$ and $\ket{\zeta_{1-}}$ as special cases. 

With a single laser field resonant to the Rydberg transition, i.e. $\Delta_1=0$, we immediately recognize 
enhanced Rabi oscillations in Eq.~\eqref{eq:effHres}, because
\begin{align} \label{eq:enhanced_Rabi}
H^{0\leftrightarrow 1}_N(\vec{k}_1)/\hbar =
\Omega_1^* \sqrt{N}\ket{g\ldots g} 
\bra{D^N_1(\vec{k}_1)} + \textrm{H.c.} 
\end{align} 
describes the well-known effect of collective Rabi enhancement, with $\Omega_N=\sqrt{N}\Omega_1$~\cite{Morgado+Whitlock:21}. 
This effect was first experimentally observed in Ref.~\cite{Gaetan+:09} and more recently discussed, for example, in Ref.~\cite{Schlosser+:20}.
Due to the site-dependent phases the oscillations appear between the ground state and the twisted $W$ state $\ket{D^N_1(\vec{k}_1)}$. 

\begin{figure}[t!]
\includegraphics[width=0.4\textwidth]{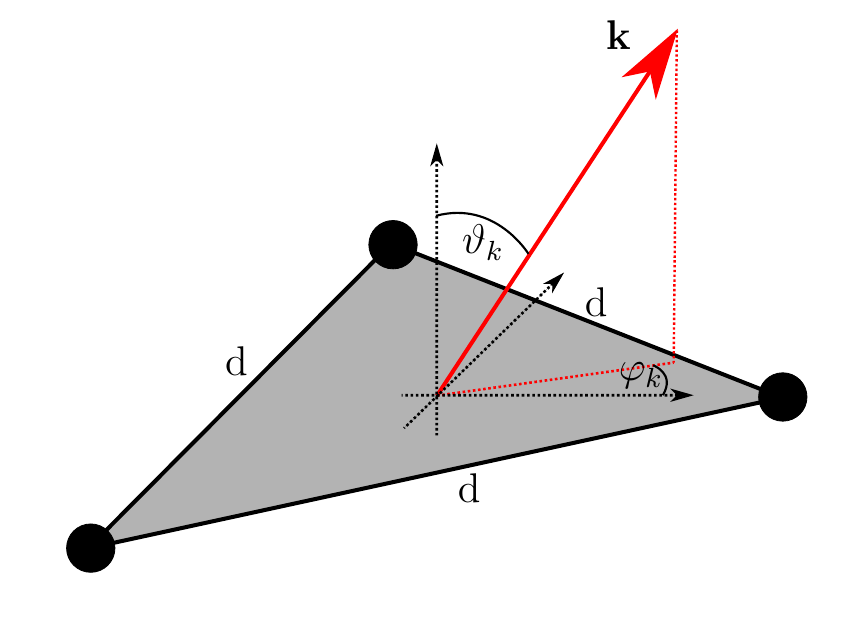}
\caption{\label{fig:prepWscheme} Schematic drawing of the orientation of the laser field and the atomic plane for 
the $N=3$ case. The atoms form an equilateral triangle with interatomic distance $d$. The orientation of a plane-wave 
laser field with wave vector $\vec{k}$ is defined through the angles $\theta_k$ and $\phi_k$.}
\end{figure}

Regarding the total number of excitations, the different twisted $W$ states are indistinguishable.
With the atomic ground state $\ket{g\ldots g}$ as initial state, and assuming control of the alignment of the 
resonant laser field, it is straightforward to prepare a specific class of twisted $W$ states by applying a laser 
pulse such that half a Rabi oscillation with Rabi frequency $\Omega_N$ is carried out.
We explicitly determine the phases in the Rydberg-trimer case ($N=3$), because it is straightforward to implement 
them on purely geometrical grounds.

If the three atoms are positioned such that they form an equilateral triangle of interatomic distance $d$, their positions 
relative to the center of mass can be described via the radial coordinate $r=d/\sqrt{3}$ and three azimuthal angles with 
relations $\phi_1-\phi_2=4\pi/3$ and $\phi_3-\phi_2=2\pi/3$.
The site-dependent phases of the laser field are given by
\begin{equation}
\vec{k}_1\cdot \vec{x}_n = \frac{\omega_A}{c}\frac{d}{\sqrt{3}} 
\sin \theta_k \cos(\phi_{k}-\phi_n) \:,
\end{equation}
where $\theta_k$ is the angle between the propagation direction of the laser field and the atomic plane, while $\phi_k$ is 
the azimuthal angle describing the projection onto this plane [cf. Fig.~\ref{fig:prepWscheme}].
Because we treat the interaction in the RWA, the resonance frequency is always much higher than the absolute values of 
the detunings, i.e. $|\Delta_j|/\omega_A\ll 1$, thus the phases are solely determined by the propagation direction, i.e. 
$\theta_k$ and $\phi_k$ and the interatomic distance $d$ (for a schematic illustration of the laser orientation with 
respect to the atomic plane, see Fig.\ref{fig:prepWscheme}).

\begin{figure}[t]
\centering
\includegraphics[width=\linewidth]{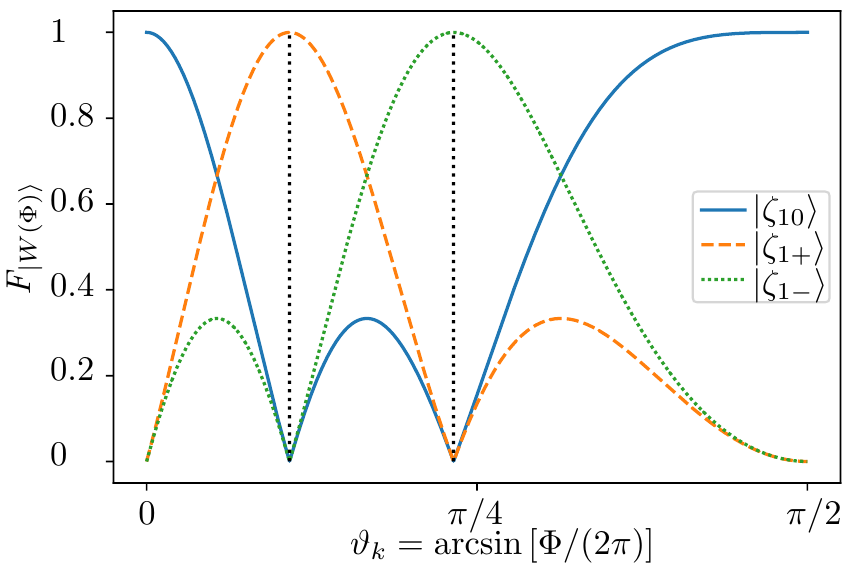}
\caption{\label{fig:prepWFid} Fidelity $\text{F}_{\ket{W(\Phi)}}=|\braket{W(\Phi)}{\zeta_{1s}}|$ 
($s=0,\pm$) of the twisted $W$ state $\ket{W(\Phi)}$ corresponding to the chiral basis states 
$\ket{\zeta_{10}}=\ket{W_3(\vec{k}=0)}$ (solid line), $\ket{\zeta_{1+}}$ (dashed line) 
and $\ket{\zeta_{1-}}$ (dotted line) as the target states, for different polar angles $\theta_k=\arcsin(\Phi/2\pi)$ 
and azimuthal angle $\phi_k-\phi_2=\pi/2$. The vertical dotted lines indicate the configurations required for 
the preparation of the two chiral states.} 
\end{figure}

\begin{figure}[b]
\centering
\includegraphics[width=\linewidth]{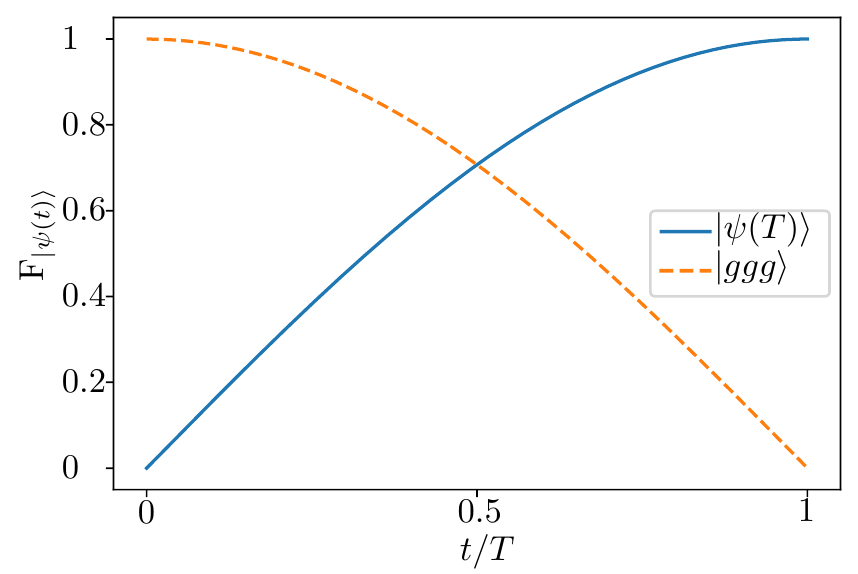}
\caption{Typical time evolution of fidelities $\text{F}_{\ket{\psi(t)}}=|\braket{\psi(t)}{\Psi}|$ 
($\ket{\Psi}=\ket{ggg},\ket{\psi(T)}$) corresponding to a $\pi$-pulse with $\Omega_1 = \pi/(2\sqrt{3}T)$, 
where $T$ is the conversion time and $\ket{\psi(T)} = \ket{W_3(\vec{k}_1)}$ the 
resulting twisted $W$ state. The time evolution corresponds to the $N=3$ Hamiltonian (cf. Eq.~\eqref{eq:Ht3fields} in
Sec.~\ref{ssec:simplifiedWtoGHZ} below) and randomly chosen phases $\vec{k}_1\cdot\vec{x}_n$, with $V/\hbar=3000/T$ and 
$\Omega_2=\Omega_3=0$.}
\label{fig:fid_pi}
\end{figure}

Since $\sum_n \cos(\phi_k-\phi_n)=0$, we can only describe symmetric twisting in this setup where $\sum_n \vec{k}\cdot\vec{x}_n=0$.
For example, we can choose the interatomic distance to be twice the resonance wavelength (i.e. $d=4\pi c/\omega_A$) and the 
relative polar angle $\phi_k-\phi_2=\pi/2$, such that $\cos(\phi_k-\phi_n)=-\sqrt{3}/2,0,\sqrt{3}/2$.
With this setup the whole range of relative phases $\vec{k}\cdot(\vec{x}_{1,3}-\vec{x}_2)=\mp\Phi$ with $0\leq \Phi \leq 2\pi$ 
is achievable by tilting the laser field accordingly with respect to the atomic plane such that $0\leq\theta_k\leq \pi/2$ and 
$\theta_k=\arcsin\left[\Phi/(2\pi)\right]$. In the envisioned scheme, half a Rabi oscillation drives the ground state $\ket{ggg}$ 
into the symmetrically twisted $W$ state
\begin{equation}\label{WPhiExpr}
\ket{W(\Phi)} = \frac{1}{\sqrt{3}}\left(e^{-i \Phi}\ket{rgg}+\ket{grg}
+e^{i\Phi}\ket{ggr}\right) \:.
\end{equation}
In particular, the two chiral states $\ket{\zeta_{1+}}$ and $\ket{\zeta_{1-}}$ can be realized by tilting the 
propagation direction of the laser field such that $\theta_k=\arcsin(1/3)$ and $\theta_k=\arcsin(2/3)$, respectively 
(note that the corresponding values of $\Phi$ are $2\pi/3$ and $-2\pi/3$). This is illustrated in Fig.~\ref{fig:prepWFid}, 
which shows the fidelities $|\braket{W(\Phi)}{\zeta_{1s}}|$ of the state $\ket{W(\Phi)}$ corresponding 
to the ordinary $W$ state ($s=0$) and the two chiral states $(s = \pm)$ dependent on the polar angle $\theta_k\equiv
\arcsin\left[\Phi/(2\pi)\right]$ of the laser field. Figure~\ref{fig:fid_pi} shows an example of a time evolution for 
the preparation of a twisted $W$-state from the ground state $\ket{ggg}$ via a $\pi$-pulse.

The fact that the external field only allows the generation of a $W$ state with one specific twisted relative 
phase [represented by $\Phi$ in Eq.~\eqref{WPhiExpr}] can be seen as a selection rule. Namely, this twisted phase has 
to match the one characterizing the field itself. In other words, the field only connects the ground state -- for which 
the analog of this twisted phase is zero -- to one particular (field-specific) $W$ state.

Given that the states $|\zeta_{10}\rangle\equiv|W_3\rangle$, $|\zeta_{1+}\rangle$, and $|\zeta_{1-}\rangle$
form an orthonormal basis of the $a=1$ subspace of the total three-atom (qubit) Hilbert space, a combination of three laser 
fields with these specific alignments could drive the ground state towards an arbitrary linear superposition of these states.
In other words, with three laser fields the preparation of an arbitrary twisted $W$ state of three qubits is possible, 
since all three orthogonal transitions are addressed.

One specific application of chiral $W$ states in QIP pertains to implementing noiseless-subsystem (NSS) qubit encoding~\cite{Knill+:00}.
NSS encoding is one of the well-known encoding schemes for logical qubits that are inherently robust to noise and 
constitute an alternative to active error correction. This type of encoding represents a three-qubit generalization of 
a two-qubit decoherence-free subspace (DFS) encoding~\cite{Lidar+:98}. While the latter is robust against global $\sigma_z$ 
dephasing, NSS encoding is insensitive to any global Pauli operator~\cite{Knill+:00}. In particular, a dissipative preparation 
of chiral $W$ states in a trapped-ion system, along with the implementation of noiseless-subsystem encoding, has quite 
recently been reported~\cite{Cole+:21}. On the other hand, the preparation of such states and the implementation of NSS 
encoding with neutral atoms in Rydberg states has never been reported before, thus the scheme proposed here may serve as 
the basis for an experimental realization.

\section{Conversions from twisted $W$ to GHZ states}\label{sec:StateEngineer}
Having considered the generation of special types of twisted $W$ states in the system at hand,
we now turn our attention to the conversion of $W$ states into their GHZ counterparts. While the creation of $W$-type
states, characterized by a single excitation that is shared by all the atoms in an ensemble, represents the hallmark 
of the RB regime~\cite{Morgado+Whitlock:21}, any realization of a GHZ state with strongly-interacting Rydberg atoms 
can be viewed as an antiblockade-type phenomenon~\cite{Su++:20}. While Rydberg antiblockade for two or more atoms 
can result from different scenarios, in the strongly-interacting regime ($| V| \geq 10\: \hbar | \Omega |$, where $\Omega$ is the 
relevant Rabi frequency of the external laser) it entails a dispersive interaction with the specific value of the 
detuning $\Delta$ of the external laser from the relevant internal transition ($\hbar \Delta=V/2$ in the two-atom case;
$\hbar \Delta=(N-1)V/N$ in the general case~\cite{Su++:20}). Our scheme for realizing GHZ states, which involves multiple 
lasers with differing detunings, is far more complicated than this conventional scenario. Yet, because it results 
in a finite probabality to have a state with more than one atom simultaneously excited to the Rydberg state in the 
strongly interacting regime it can be considered as a generalized form of the Rydberg antiblockade.

The structure of the ladder Hamiltonian in Eq.~\eqref{eq:effHLadder} is such that it only connects adjacent 
energy levels of the atomic ensemble. However, due to the existence of nontrivial offsets, the dynamics inherent 
to this Hamiltonian are not necessarily enclosed within a subspace of $N+1$ states. In the following, we discuss two 
approaches whereby one can ensure subspace-enclosed dynamics by selecting one state for each excitation number $a$.
In these cases the system can be described by an effective Hamiltonian connecting adjacent levels, thus inheriting 
the already existing solutions for systems described by Hamiltonians of that type. In all the following cases this is 
accomplished through a readjustment of Rabi frequencies. In what follows, we will either make use of a simple
$\pi$ pulse to drive half a Rabi oscillation, or an adaption of a more complicated pulse scheme that was utilized
for conversions between $W$ and GHZ states in Ref.~\cite{Haase+:21}.

In Sec.~\ref{ssec:simplifiedWtoGHZ} below, we discuss a scheme where the alignments of the laser fields and the positioning 
of the atoms are adjusted to ensure a subspace-enclosed dynamics of $N+1$ states via selection rules as in the state-preparation 
scheme of Sec.~\ref{sec:preptwistedW}. By contrast, in Sec.~\ref{ssec:twistcompensation} we 
consider an alternative scheme in which an additional laser field lifts some of the degeneracies and the states that participate 
in the dynamics are singled out via fine detunings of the remaining laser fields and the attendant hierarchies of timescales. 
This last scheme is then discussed in a broader context in Sec.~\ref{TimeHierarch}, where we also demonstrate its soundness 
by showing that typical state-conversion times are much shorter than the relevant Rydberg-state lifetimes.

\subsection{Conversion schemes involving degenerate Dicke manifolds of states} \label{ssec:simplifiedWtoGHZ}
In contrast to the preparation scheme in Sec.~\ref{sec:preptwistedW}, the presence of several laser fields 
complicates the situation as they have, in principle, different propagation directions (i.e. different wave 
vectors $\vec{k}_j$). Equation~\eqref{eq:effHLadder} can in this case be written as
\begin{multline} \label{eq:effHladder_again}
U^\dagger(\vec{k}_1)H^\text{L}_N(\{\vec{k}_a\})U(\vec{k}_1) \\
= \sum_{a=1}^N \hbar\Omega^*_a 
U(\vec{k}_a-\vec{k}_1)\sigma_a^-U^\dagger
(\vec{k}_a-\vec{k}_1) +\textrm{H.c.} \: .
\end{multline}
The alignment of the laser field resonant to the Rydberg transition ($\Delta_1=0$) sets a reference frame insofar 
that it is the only one connecting the remaining levels to the ground state.
Any state conversion scheme naturally starts with all atoms in the ground state or is preceded by a preparation 
scheme of the kind proposed in Sec.~\ref{sec:preptwistedW}.
Because of that and without loss of generality, we set the twisting induced by the first laser as the reference 
one, i.e. we set $\vec{k}_1\cdot\vec{x}_n=0\mod 2\pi$.
This transforms all wave vectors $\vec{k}_j$ to $\vec{k}_j-\vec{k}_1$.
If all laser fields are properly aligned such that \mbox{$(\vec{k}_j-\vec{k}_1)\cdot\vec{x}_n=0 \mod 2\pi$}, 
we can split up the $a$-th rising operator into a parallel and orthogonal part with respect to 
$\ket{D^N_a}$ such that the dynamics of $\ket{D^N_a}$ states decouples from their orthogonal counterparts:
\begin{align} \label{eq:effHladder_simpl}
H^\text{L}_N/\hbar =  \sum_{a=1}^N \Omega_a^* \Big[&\sqrt{a(N-a+1)}\ket{D^N_{a-1}}\bra{D^N_a} \nonumber \\
& +\sigma_a^-(P_a-\ket{D^N_a}\bra{D^N_a})\Big] +\textrm{H.c.}
\end{align}
Now with $P_D = \sum_{a=0}^N \ket{D^N_a}\bra{D^N_a}$ we project onto the subspace just containing the 
$N+1$ different $\ket{D^N_a}$ states. Hence, we calculate $P_D H_N^\text{L}P_D$ and get
\begin{align}\label{exprHdl}
H^\text{DL}_N/\hbar=  \sum_{a=1}^N \Omega^*_a 
\sqrt{a(N-a+1)}\ket{D^N_{a-1}}\bra{D^N_a} + \textrm{H.c.} \: .
\end{align}
The resulting effective Hamiltonian is a matching ladder of Dicke states (DL), hence it connects stepwise all
$N+1$ energy levels, such that any state conversion involving adjacent energy levels that are connected via Rabi 
frequencies can be carried out. For example, pulses not overlapping in time, which induce
Rabi half-oscillations corresponding to adjacent transitions would drive the system from the ground- 
to the highest excited state. Alternative schemes with temporally-overlapping pulses are also possible.

We now discuss some special cases of Eq.~\eqref{exprHdl}. For $N=3$ we obtain
\begin{align}
H^\text{DL}_3/\hbar = {} &  \sqrt{3}\Omega^*_1 \ket{ggg}\bra{D^3_1} + 2 \Omega^*_2 \ket{D^3_1}\bra{D^3_2} \nonumber \\ 
						 & + \sqrt{3}\Omega^*_3 \ket{D^3_2}\bra{rrr} +\textrm{H.c.} \: .
\end{align}
Similarly, for $N=4$ we have
\begin{align}
H^\text{DL}_4  /\hbar = {} & 2\Omega^*_1 \ket{gggg}\bra{D^4_1}+ \sqrt{6} \Omega^*_2 \ket{D^4_1}\bra{D^4_2} \nonumber \\ 
&+ \sqrt{6}\Omega^*_3\ket{D^4_2}\bra{D^4_3} 
+ 2\Omega^*_4 \ket{D^4_3}\bra{rrrr} +\textrm{H.c.} \: .
\end{align}
These are the same effective Hamiltonians as used in \cite{Zheng+:20,Haase+:21} for $W$-to-GHZ state conversion.
However, it is important to point out that a strong off-resonant laser field, as it was utilized in this previous studies, 
is not a prerequisite for obtaining these effective Hamiltonians, as long as the state of the atomic ensemble fulfills 
$P_D \ket{\psi(t)}=\ket{\psi(t)}$ for all times $t$ during the conversion process.
This can be achieved by properly aligning all laser fields as discussed above. With other alignments, effective 
Hamiltonians which include orthogonal chiral states can be designed exploiting selection rules and the additional 
twisting induced by the laser fields. We will carry this out explicitly for the Rydberg-trimer case ($N=3$).

\begin{figure}
\includegraphics[width=\linewidth]{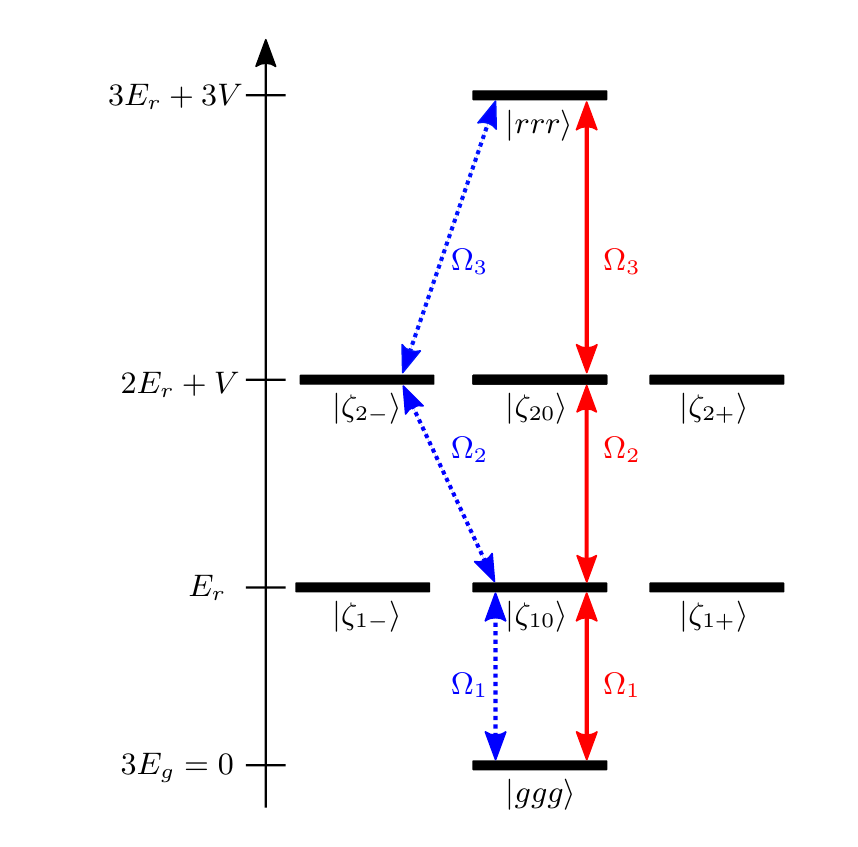}
\caption{Energies of the atomic ensemble with $N=3$ corresponding to the states $\ket{\zeta_{as}}$ [cf. Eqs.~\eqref{Wlike}].
The solid arrows indicate transitions driven in the case when all laser fields are aligned such that $(\vec{k}_j-\vec{k}_1) 
\cdot\vec{x}_n=0$, while the dotted ones indicate those corresponding to the choice of laser-field alignments described by 
the upper-sign case of Eq.~\eqref{eq:alignments}. Here $E_g=0$ is chosen as origin for the energy scale.}
\label{fig:levelscheme_chiral}
\end{figure}

In the basis of $\ket{\zeta_{as}}$ states [cf. Eqs.~\eqref{Wlike}], we can recast Eq.~\eqref{eq:effHladder_again} in the form
\begin{align} \label{eq:effHladder_3}
H^\text{L}_3(\{\vec{k}_a\})/\hbar 
= {} & \Omega_1^* \sqrt{3}\:\ket{ggg}\bra{\zeta_{10}} \nonumber \\ 
 	 &+\Omega^*_3 \sqrt{3}U(\vec{k}_3)\ket{\zeta_{20}}\bra{rrr}U^\dagger(\vec{k}_3)
\nonumber \\ 
	 &+\Omega^*_2\:U(\vec{k}_2)\Big[2\ket{\zeta_{10}}\bra{\zeta_{20}}-\ket{\zeta_{1+}}\bra{\zeta_{2+}}
 \nonumber \\ &
 -\ket{\zeta_{1-}}\bra{\zeta_{2-}}\Big]U^\dagger(\vec{k}_2) +\textrm{H.c.} \:. 
\end{align}
Atom positions and laser-field alignment chosen such that
\begin{align}\label{eq:alignments}
U(\vec{k}_2)\ket{\zeta_{1\pm}} = {} & \ket{\zeta_{10}}  \, ,&  U(\vec{k}_2)\ket{\zeta_{2\pm}}
= {} & U(\vec{k}_3)\ket{\zeta_{20}}
\end{align}
would single out a $-\ket{\zeta_{10}}\bra{\zeta_{2-}}$ ($-\ket{\zeta_{10}}\bra{\zeta_{2+}}$) transition operator 
in the Hamiltonian in the upper-sign (lower-sign) case.
If the initial state lies in the subspace spanned by the four states $\{\ket{ggg},\ket{W},\ket{\zeta_{2-}}(\ket{\zeta_{2+}}),
\ket{rrr}\}$ the unitary time evolution of the atomic ensemble is enclosed in this subspace.
State conversion schemes where the underlying Hamiltonian connects adjacent levels can easily adapted by adjusting the Rabi frequencies.
Effective Hamiltonians including different states of the $a=2$ subspace are indicated in the level scheme of Fig.~\ref{fig:levelscheme_chiral}.

To illustrate the differences in the effective Hamiltonian with respect to the alignments the evolution of the initial 
state $\ket{\psi(t=0)}=\ket{W}$ is calculated numerically~\cite{Johansson+:12,Johansson+:13,Hunter:07,Harris:20} based 
on the interaction Hamiltonian
\begin{align} \label{eq:Ht3fields}
H_I(t) = {} & \sum_{j=1}^3\sum_{n=1}^3\left(\hbar\Omega_{i}e^{i(\vec{k}_j\cdot\vec{x}_n-
\Delta_j t)}\ket{r}_{nn}\bra{g} + \textrm{H.c.}\right) \nonumber \\
 & +\sum_{p<q} V_{pq}\ket{rr}_{pq}\bra{rr}
\end{align}
with constant Rabi frequencies $\Omega_i$ realizing $W$ to GHZ conversion via $\ket{\zeta_{20}}$.

In the case when all laser fields are aligned, the values of the Rabi frequencies are $T_0\Omega_1=1.22/\sqrt{3}$, 
$T_0\Omega_2 = 1.42/2$ and $T_0\Omega_3=2.35/\sqrt{3}$, where $T_0$ is the conversion time.
These specific values of the constant Rabi frequencies are determined in Ref.~\cite{Haase+:21} and are based on the observation,
that under the assumption of real-valued Rabi frequencies state conversion in a four-level system is characterized by the 
dynamical symmetry $\text{su}(2)\oplus\text{su}(2)\cong \text{so}(4)$. 
Therefore, it can effectively be described in the form of two pseudospin-$1/2$ degrees of freedom. 
The fact that only terms connecting adjacent excitation subspaces ($a-1\leftrightarrow a \leftrightarrow a+1$) appear in the 
effective Hamiltonian introduces constraints to the full dynamics of the two pseudospins.

For the conversion between $W$ and GHZ states via $\ket{\zeta_{2-}}$ [cf. the upper-sign case of Eq.~\eqref{eq:alignments}] 
to be carried out in the same time $T_0$, the second Rabi frequency would have to be doubled. In order to be able to compare  
the two conversion paths, we adjust all Rabi frequencies such that the total laser-pulse energy over the corresponding 
conversion time is the same in both cases.
The total laser-pulse energy is given by
\begin{equation} \label{eq:laserenergy}
A(t)=\int_0^{t}\sum_{j=1}^3|\Omega_j(t')|^2 \text{d}t' \:.
\end{equation}
(The time dependence of the Rabi frequencies is introduced here only for later convenience.)
Both schemes allow one to carry out the desired state conversion, but the conversion via the achiral state is faster under 
the assumption of equal laser-pulse energy consumption.
In both schemes only one of the $a=2$ states acts as intermediate state in the conversion process while the other such states 
are never occupied. The target state in both cases is
\begin{equation} \label{eq:GHZ}
\ket{\textrm{GHZ}} = \frac{1}{2}\left(\ket{ggg}+e^{-i 3 V t}\ket{rrr}\right) \:,
\end{equation}
where the time-dependent relative phases account for the energy shift arising due to the constant energy difference 
between levels in Eq.~\eqref{eq:Ht3fields}. 

\begin{figure}
\includegraphics[width=\linewidth]{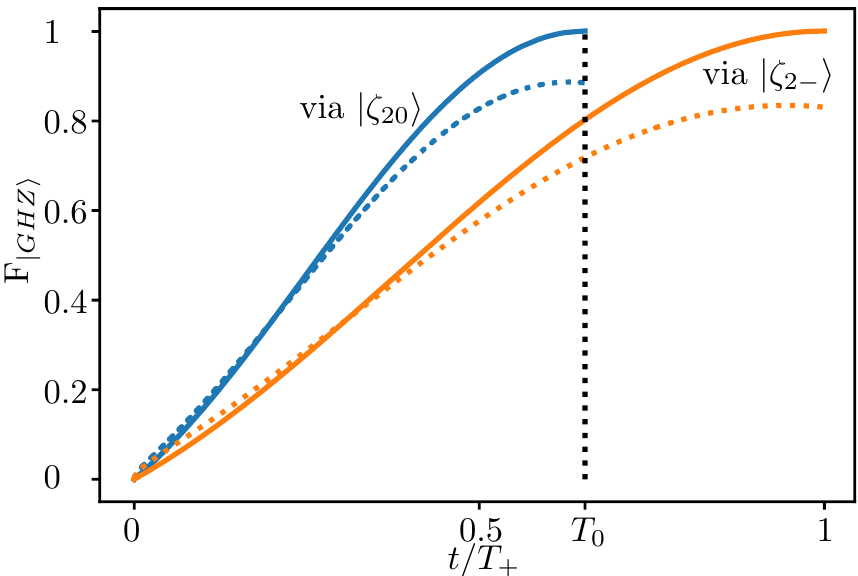}
\caption{Time dependence of target-state fidelities $\text{F}_{\ket{\textrm{GHZ}}}=\sqrt{\bra{\textrm{GHZ}}
{\rho(t)}\ket{\textrm{GHZ}}}$ [cf. Eq.~\eqref{eq:GHZ}] for the $W$-to-GHZ state conversions via two different intermediate 
states $\ket{\zeta_{20}}$ and $\ket{\zeta_{2,-}}$, both shown for $VT_+/\hbar=3000$. The solid lines correspond to the unitary 
dynamics where $\rho(t)=\ket{\psi(t)}\bra{\psi(t)}$. The dotted lines correspond to the open-system dynamics 
with the dephasing- and spontaneous-decay rates $\Gamma=\gamma=0.1/T_+$. The two conversion pathways are adjusted such 
that their respective total laser-pulse energy consumptions are mutually equal.}
\label{fig:conversionschemes_target}
\end{figure}

The results obtained in numerical calculations, which correspond to $V_{pq}=V$ and $V/\hbar=3000/T_+$ in both cases (where 
$T_+$ is the conversion time in the upper-sign case), are shown in Fig.~\ref{fig:conversionschemes_target}. 
What can be inferred from these results is that -- while both conversion schemes realize the target state -- the scheme 
that makes use of $\ket{\zeta_{20}}$ as intermediate state requires a significantly shorter time than the one where 
$\ket{\zeta_{2-}}$ plays the analogous role. For the sake of completeness, it is should be stressed that yet another 
state-conversion pathway -- equivalent to the second one -- that makes use of $\ket{\zeta_{2+}}$ as its intermediate state,
is also possible [lower-sign case of Eq.~\eqref{eq:alignments}]. 

In realistic experimental setups spontaneous decay of the Rydberg state and dephasing, caused e.g. by atomic 
motion~\cite{Shi:20}, limit the lifetime and the accuracy of the proposed state-conversion schemes.
To take such effects into account, we characterize the corresponding open-system dynamics within the framework 
of the Lindblad master equation~\cite{BreuerPetruccioneBOOK:02}.
In this framework, the dynamics of the density operator $\rho(t)$ is governed by the equation
\begin{align}
\frac{d\rho}{dt} =& -\frac{i}{\hbar}[H_I(t),\rho(t)] \nonumber \\
&+ \sum_{l=1}^2 \frac{1}{2}
\left([\rho(t)L_l,L_l^\dagger]+[L_l\rho(t),L_l^\dagger]\right) \:,
\end{align}
where the two relevant Lindblad operators are given by
\begin{equation} \begin{split}
L_1 = {} & \sqrt{\Gamma}\sum_{n=1}^3 \ket{g}_{nn}\bra{r} \: , \\
L_2 = {} & \sqrt{\gamma} \sum_{n=1}^3(\ket{g}_{nn}\bra{g}-\ket{r}_{nn}\bra{r}) \:. 
\end{split} \end{equation} 
Here $L_1$ describes spontaneous decay from the Rydberg- to the ground state of an atom 
with decay rate $\Gamma$, while $L_2$ describes the dephasing of these states with the rate $\gamma$.
We solved the last Lindblad master equation numerically~\cite{Johansson+:12,Johansson+:13,Hunter:07,Harris:20}, 
choosing rather high rates $\Gamma=\gamma=0.1/T_+$. Needless to say, the target-state fidelity 
[cf. Fig.~\ref{fig:conversionschemes_target}] in the open-system scenario is smaller than those found 
in the closed-system treatment. The obtained results for the fidelity speak in favor of using the faster 
conversion path, as the debilitating effects of spontaneous decay and dephasing are weaker for that path.

The preparation of chiral states discussed in Sec.~\ref{sec:preptwistedW} and the state-conversion scheme presented here, 
rely heavily on setting the site-dependent phases $\vec{k}_j\cdot\vec{x}_n$. Therefore, it is necessary to control the 
orientation of each laser field relative to the atomic plane and the position of the atoms to a precision of the order 
of the laser wavelength. Owing to the recent advances in manipulation and control of cold neutral atoms in optical 
tweezers~\cite{Norcia+:18,Liu+:19,Deist+:22}, this last requirement is within experimental reach.

To describe the influence of fluctuations in the atomic positions on the conversion scheme, we consider 
random variations of atomic positions. These variations affect not only the phases $\vec{k}_j\cdot\vec{x}_n$, but also the 
interatomic potentials $V_{pq}$, because they cause the arrangement of atoms to deviate from the original equilateral triangle.
To differentiate these two effects, it is prudent to concentrate on the faster conversion scheme via $\ket{\zeta_{20}}$.
Because in this case all laser fields are aligned, we have $\vec{k}_j\cdot{\vec{x}_n}=\vec{k}_{j'}\cdot\vec{x}_n$. As already 
discussed above, $\vec{k}_1\cdot \vec{x}_n$ sets the reference phase. Hence, if we neglect a misalignment of laser fields, 
the conversion scheme via $\ket{\zeta_{20}}$ is not affected by the random phases. No further matching conditions as in 
Eq.~\eqref{eq:alignments} have to be fulfilled. Furthermore, due to the scaling of $V_{pq} = C_6/d_{pq}^6$, the influence of 
varying interaction potentials $V_{pq}$ can be expected to dominate over small variations in the phase-matching conditions.

In order to quantify the effect of the deviation from its original value $V=C_6/d^6$ at interatomic distance $d$, we 
computed the different interatomic potentials $V_{pq}= V d^6/d^6_{pq}$ according to randomly sampled atomic positions. We introduce 
random errors for each of the spatial coordinates of the three atoms. Accordingly, $\vec{x}_n \to \vec{x}_n + \vec{\epsilon}_n$ 
differs for all atoms $n$ ($n=1,2,3$). In each realization, the nine components of the three error vectors $\vec{\epsilon}_n$ were 
independently drawn from a standard normal distribution of standard deviation $\sigma$ resulting in varying distances $d_{pq}$ and, 
accordingly, three different $V_{pq}$ per realization. We then numerically computed the time evolution according to Eq.~\eqref{eq:Ht3fields}, 
where random positioning error vectors $\vec{\epsilon}_n$ were drawn componentwise from a standard normal distribution with 
$\sigma \in [0,0.1\lambda_0]$, where $\lambda_0$ is the resonance wavelength, and a sample size $S=500$. This numerical 
evaluation was repeated for different choices of the standard deviation $\sigma$ resulting in different standard deviations 
$\sigma_d=\sqrt{\overline{(d_{pq}-\overline{d}_{pq})^2}}$ of all $3S$ different values of $d_{pq}$ per sample 
(where $\overline{d}_{pq}$ is the mean of all $3S$ $d_{pq}$ per sample). The parameter values used in these calculations 
were $V/\hbar=30.86/T_0$, where $T_0$ is the conversion time, and $d=40\lambda_0$.

\begin{figure}
\includegraphics[width=\linewidth]{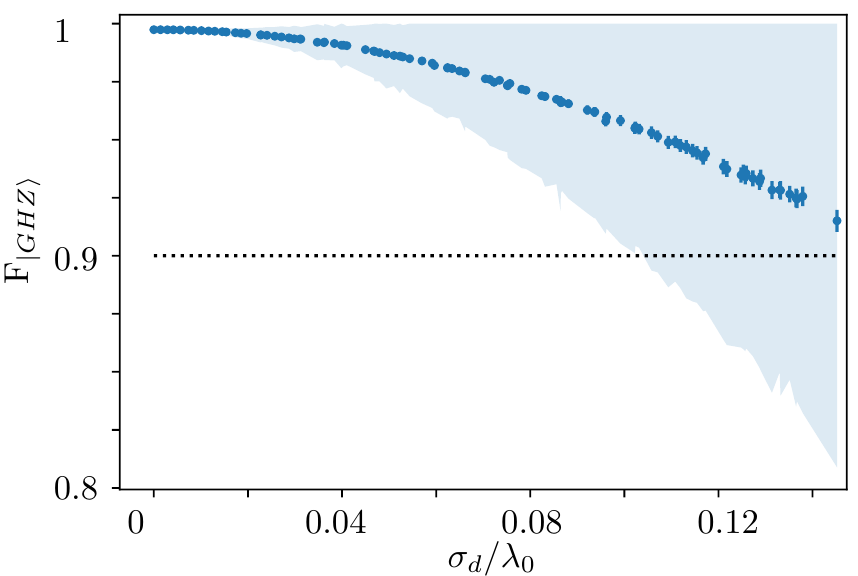}
\caption{Mean values of the GHZ-state fidelity $\text{F}_{\ket{\textrm{GHZ}}}$ and their corresponding 
standard deviations $\sigma_\text{F}$ (shaded area) corresponding to the $W$-to-GHZ state conversion via $\ket{\zeta_{20}}$, 
computed from a sample of $S=500$ results, for varying standard deviation $\sigma_d$ of the interatomic distance. The error 
bars show the standard error $\sigma_\text{F}/\sqrt{S}$ of the mean fidelity. The parameter values used are $V/\hbar=30.86/T_0$ 
and $d=40\lambda_0$, where $T_0$ is the state-conversion time and $\lambda_0$ the resonance wavelength.}
\label{fig:mean_fidelities_eror}
\end{figure}

Figure~\ref{fig:mean_fidelities_eror} shows the obtained mean values of the GHZ-state fidelity
$\text{F}_{\ket{\textrm{GHZ}}}=|\braket{\textrm{GHZ}}{\psi(T_0)}|$ and its corresponding standard deviation 
$\sigma_\text{F}=\sqrt{\overline{(\text{F}_{\ket{\textrm{GHZ}}}-\overline{\text{F}_{\ket{\textrm{GHZ}}}})^2}}$
for different values of $\sigma_d$ at $t=T_0$ for the $W$-to-GHZ state conversion via $\ket{\zeta_{20}}$. It can 
be inferred from the obtained results that the mean values of the fidelity are above $0.9$ for the whole range of 
considered values of $\sigma$. The chosen simulation parameters were assumed to have values characteristic 
of alkali atoms most often used in optical-tweezer experiments, with the principal quantum number $n=50$ and the 
interatomic distance $d=4\:\mu$m~\cite{Adams+:20}. This speaks in favor of the experimental feasibility of the 
proposed state-conversion scheme.

In addition to the already presented conversion scheme, we discuss an alternative approach in 
Sec.~\ref{ssec:twistcompensation}. That approach makes use of an additional laser field to set energy shifts in 
the atomic Hamiltonian. This singles out a specific twisted $W$ state determined by the site-dependent phases of 
this strong driving field. In turn, this allows one to address specific atomic states via fine detunings $\delta_j$ 
of the other laser fields ($j=1,2,3$).

\subsection{Conversion schemes involving lifted degeneracies} \label{ssec:twistcompensation}
The effective Hamiltonians derived for state-conversion tasks in Sec.~\ref{ssec:simplifiedWtoGHZ} depend on proper 
relative alignment of the resonant laser fields involved and precise positioning of the Rydberg atoms.
Misalignment or errors in the positioning of the atoms result in unwanted phase shifts.
We can use a combination of nearly resonant fields (enumerated by $j=1,2,3$) and an additional stronger field $(j=0)$, 
where the latter sets energy shifts such that it lifts some of the degeneracies of $H_A$.
Fine detunings $\delta_i$ added onto the detunings $\Delta_i$ can then address specific transitions.
This procedure is inspired by the derivation of the effective Hamiltonian in Ref.~\cite{Zheng+:20} 
but realizes a generalized version including relative twisting.

Again, we explicitly calculate it for the Rydberg trimer case $N=3$.
The corrections to $H_A$ in this case are given by
\begin{equation}
H_3^\text{shift} = H^\text{off}_3(\vec{k}_0) + H^\text{L}_3(\{\vec{k}_a\}) \: .
\end{equation}
It is the combination of the off-resonant case from Eq.~\eqref{eq:effH_offres3} and the ladder Hamiltonian 
Eq.~\eqref{eq:effHLadder} for $N=3$ being
\begin{equation} \begin{split}
H^\text{L}_3 (\{\vec{k}_a\})/\hbar = \bigg[&\Omega_1^*\sqrt{3}\ket{ggg}\bra{D^3_1(\vec{k}_1)} \\
&+\Omega_2^*\Big(3\ket{D^3_1(\vec{k}_2)}\bra{D^3_2(\vec{k}_2)}
\\
&-\Phi^*(\vec{k}_2)%e^{-i\vec{k}_2\cdot(\vec{x}_1+\vec{x}_2+\vec{x}_3)}
\sum_{n=1}^3e^{2i\vec{k}_2\cdot\vec{x}_n} \ket{gg}\ket{r}_{nn}\bra{g}\bra{rr}\Big)
 \\
&+\sqrt{3}\Omega_3^* \ket{D^3_2(\vec{k}_3)}\bra{rrr}
\Phi^*(\vec{k}_3)%e^{-i\vec{k}_3\cdot(\vec{x}_1+\vec{x}_2+\vec{x}_3)}
\bigg] + \textrm{H.c.} 
\end{split}\end{equation}
with $\Phi(\vec{k})=e^{i\vec{k}\cdot\sum_n^N \vec{x}_n}$.
The overlaps of twisted states corresponding to different fields are 
\begin{equation} \begin{split}
\braket{D^3_1(\vec{k}_1)}{D^3_1(\vec{k}_{2})} = {}& \frac{1}{3} \Sigma_{\vec{k}_2-\vec{k}_1} \\
\braket{D^3_2(\vec{k}_2)}{D^3_2(\vec{k}_{3})} = {}& \frac{e^{i(\vec{k}_3-\vec{k}_2)\cdot\sum_{n=1}^3 \vec{x}_n}}{3} 
\Sigma_{\vec{k}_3-\vec{k}_2}\ 
\end{split} \end{equation}
with $0\leq\Sigma_{\vec{k}}=\sum_{n=1}^3 e^{i\vec{k}\cdot \vec{x}_n}\leq 3$, which describes the amount of 
relative twisting between two different twisted states of the same excitation number.
If there are no relative phase differences, $\Sigma_{\vec{k}_j-\vec{k}_i}\to 3$.
If $\Sigma_{\vec{k}_j-\vec{k_i}}=0$, both laser fields address orthogonal states and the effective Hamiltonian 
would split in several unconnected dynamics. However, as long as the overlaps do not vanish we can compensate 
for it by driving transitions with higher Rabi frequencies.

To show that, we transform the effective Hamiltonian to an interaction picture with respect to the stronger 
off-resonant laser field with Rabi frequency $\Omega_0\gg \Omega_j$ ($j=1,2,3$).
Since $[H_A,H^\text{off}_3(\vec{k}_0)]=0$, $H_A$ is not affected by this transformation, and we evaluate 
the remaining parts as
\begin{multline} \label{eq:ladderHoff}
e^{iH^\text{off}_3(\vec{k}_0)t/\hbar}H^\text{L}_3(\{\vec{k}_a\})e^{-iH^\text{off}_3
(\vec{k}_0)t/\hbar}/\hbar \\
=U(\vec{k}_0)e^{iH^\text{off}_3 t/\hbar}
\sum_{a=1}^3\bigg[\Omega^*_a 
U(\vec{k}_a-\vec{k}_0)\sigma_a^- U^\dagger(\vec{k}_a-\vec{k}_0)\\
+\textrm{H.c.}\bigg]
e^{-iH^\text{off}_3t/\hbar}U^\dagger(\vec{k}_0)  \: ,
\end{multline}
where $H^\text{off}$ without wave-vector argument stands for the operator without any twisting.
Now we can introduce small fine detunings $\delta_j$ to the resonant fields such that the total detunings 
are $\Delta_j^\text{total}=\Delta_j+\delta_j$.
If $|\delta_j|\ll|\Delta_0|, V/\hbar$, the fine detunings do not change the calculation of the effective 
Hamiltonians as discussed in Sec.~\ref{ssec:resonant} and Appendix~\ref{app:effHresonant}, since $\delta_j$ 
never contributes significantly. 

However, in an interaction picture with respect to $H_0=H_A+H_F$ the Rabi frequencies are shifted $\Omega_j\to
\Omega_j e^{-i\delta_j t}$ (associated with the atomic rising operator) due to the fine detunings being part
of the time dependencies of the field operators.
Now these fine detunings can be used to compensate the oscillatory behavior of one term per laser field appearing 
in Eq.~\eqref{eq:ladderHoff}. Unwanted terms will still oscillate with different residual frequencies $\omega_R$. 
However, if $\min\{|\omega_R|\}\gg T^{-1}$, where $\{|\omega_R|\}$ is the set of all the relevant residual 
frequencies and $T$ is the conversion time in question, we can ignore all terms with non-vanishing exponents 
in Eq.~\eqref{eq:ladderHoff}. By choosing
\begin{eqnarray} \label{eq:finedetunings} 
\delta_1 &=& (-6s_0+4s_1)/\hbar \:, \nonumber\\
\delta_2 &=& (3s_0-8s_1+3s_2)/\hbar \, , \\
\delta_3 &=& (4s_1-6s_2)/\hbar \nonumber\:, 
\end{eqnarray}
with $s_a = \hbar^2| \Omega_0|^2/(\hbar\Delta_0 - aV)$ ($a=0,1,2$),
we obtain the following twisted-ladder (TL) Hamiltonian:
\begin{align} \label{eq:twistedLadder} 
\widetilde{H}^\textrm{TL}_3/\hbar =& \sqrt{3}\Omega_1 \frac{|\Sigma_{\vec{k}_0-\vec{k}_1}|}{3} 
\widetilde{\ket{ggg}}\bra{D^3_1(\vec{k}_0)} \nonumber \\
&+ \sqrt{3}\Omega_3\frac{ |\Sigma_{\vec{k}_0-\vec{k}_3}|}{3} \widetilde{\ket
	{D^3_2(\vec{k}_0)}}\widetilde{\bra{rrr}} 
\nonumber \\& + 2\Omega_2 \frac{|\Sigma_{\vec{k}_0-\vec{k}_2}|}{3} \ket{D^3_1(\vec{k}_0)}\widetilde{\bra{D^3_2(\vec{k}_0)}} 
+ \textrm{H.c.} \: . 
\end{align}
Here, in order to ensure that Rabi frequencies are real-valued, we included additional phases into the redefined 
atomic states
\begin{align} \label{eq:statestwistedladder}
\widetilde{\ket{ggg}}&=e^{i\phi(\Sigma_{\vec{k}_0-\vec{k}_1})}\ket{ggg} \:,  \nonumber  \\
 \widetilde{\ket{D^3_2(\vec{k}_0)}}&=e^{-i\phi(\Sigma_{\vec{k}_0-\vec{k}_2})}\ket{D^3_2(\vec{k}_0)} \:,\\
\widetilde{\ket{rrr}} &= e^{i\left[\phi(\Sigma_{\vec{k}_0-\vec{k}_2})-\phi(\Sigma_{\vec{k}_0-\vec{k}_3})
	+\vec{k}_0\cdot\sum_{n=1}^3 \vec{x}_n\right]}\ket{rrr} \: , \nonumber 
\end{align}
where $\phi(z)$ is the argument of the complex number $z$. The driven transitions are indicated in 
Fig.~\ref{fig:levelscheme_Hoff}. A more detailed derivation of the twisted-ladder Hamiltonian 
can be found in Appendix~\ref{app:twistcompensation}.

\begin{figure}
\includegraphics[width=\linewidth]{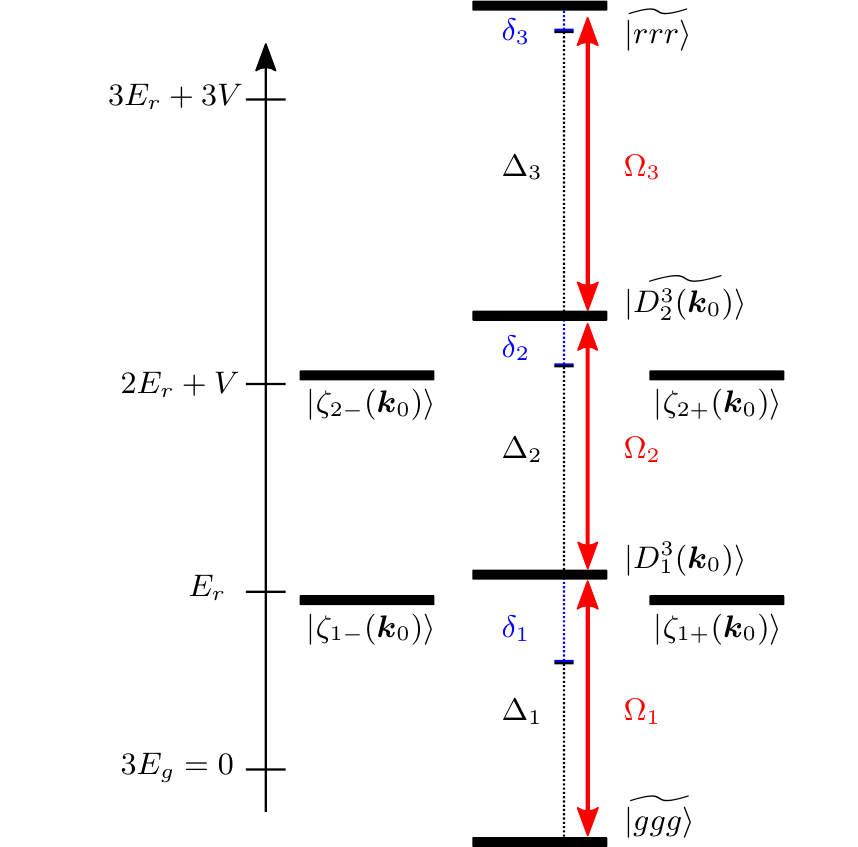}
\caption{\label{fig:levelscheme_Hoff}
Energy-level scheme corresponding to the Hamiltonian $H_A+H_3^\text{off} (\vec{k}_0)$ for Rydberg trimers.
$E_g$ is chosen as origin of the energy scale.
The arrows indicate the transitions driven by laser fields with detunings $\Delta_i+\delta_i$ as determined by 
Eq.~\eqref{eq:finedetunings} [cf. Eq.~\eqref{eq:twistedLadder}].}
\end{figure}

The result is a Hamiltonian connecting twisted states with adjacent numbers of excitations.
The twisting is solely determined by the site-dependent phases of the strong laser field ($j=0$).
Choosing other fine detunings $\delta_j$ would result in different residual frequencies and other effective 
Hamiltonians, e.g. including chiral states [relative to $U(\vec{k}_0)\ket{\zeta_{10}}$]. Yet, because the 
chiral states are still energetically degenerate with respect to $H^\text{off}_3(\vec{k}_0)$, 
the effective Hamiltonian would be of higher dimension. If all $\vec{k}_a\cdot \vec{x}_n \to 0$, i.e. without 
any twisting, this final result reproduces the effective Hamiltonian as discussed in Ref.~\cite{Zheng+:20}.
However, here we derived a generalized version which includes relative twisting due to the different laser fields.
The smaller the specific $|\Sigma_{\vec{k}_0-\vec{k}_a}|$ becomes, the higher the corresponding Rabi frequency 
$\Omega_a$ has to be for a specific conversion to be possible in a given time frame.
Those adjustments are only possible as long as all $|\Sigma_{\vec{k}_0-\vec{k}_a}|$ are not too small, because 
with increasing Rabi frequencies the perturbative treatment eventually breaks down.
Alternatively, the conversion time has to be increased accordingly which allows the Rabi frequencies to remain 
sufficiently small.

With the effective Hamiltonian~\eqref{eq:twistedLadder} we can consider the preparation of twisted $W$ states as 
in Sec.~\ref{sec:preptwistedW}, but now the amount of twisting is solely determined by the stronger off-resonant 
field ($j=0$). With $\Omega_2=\Omega_3=0$ we have an effective Hamiltonian of Eq.~\eqref{eq:twistedLadder} describing 
Rabi oscillations between $\ket{ggg}$ and $\ket{D^3_1(\vec{k}_0)}$ with effective Rabi frequency $\Omega' = \sqrt{3} \Omega_1|
\Sigma_{(\vec{k}_0-\vec{k}_1)}|/3$.
Therefore, under the assumptions of equal laser-pulse energy consumption [cf. Eq.~\eqref{eq:laserenergy}] the conversion
time increases as $T=9 T_0/|\Sigma_{\vec{k}_0-\vec{k}_1}|^2$, where $T_0$ is the reference conversion time without any 
relative twisting.
Furthermore, the constant laser fields from Ref.~\cite{Haase+:21} as already used in the last subsection implement a 
state conversion from twisted $\ket{W(\vec{k}_0)}=\ket{D^3_1(\vec{k}_0)}$ to the GHZ state
\begin{equation}
\ket{\textrm{GHZ}(\vec{k}_0)}=\frac{1}{\sqrt{2}} \left(\widetilde{\ket{ggg}}
+e^{i\phi}\widetilde{\ket{rrr}}\right) \:,
\end{equation}
if Rabi frequencies are adjusted by $3/|\Sigma_{\vec{k}_0-\vec{k}_a}|$ for $a=1,2,3$, respectively. 

To illustrate the discussed adjustments, we give an example combining preparation of the twisted state $\ket{D^3_1(\vec{k}_0)}$ 
from the ground state, followed by its conversion into a GHZ state for different amounts of relative twisting.
The laser fields $j=1,2,3$ are all aligned with the same polar angles $\theta_k$ [cf. Fig.~\ref{fig:prepWscheme}]) and 
azimuthal angle $\phi_k$. We compare three different alignments labeled via $s=3\sin(\theta_k)=0,0.5,0.75$, such that 
$|\Sigma_{\vec{k}_0-\vec{k}_j}|=3,2,1$. The first part is executed via a $\pi$-pulse of the field $j=1$ and the second 
one via the constant Rabi frequencies as mentioned before. The $\pi$-pulse is set to take a quarter of the total respective 
conversion time $T_s$. All values of Rabi frequencies ($j=1,2,3$) are adjusted such that the total laser-pulse energy consumption 
is the same in all cases [cf. Eq.~\eqref{eq:laserenergy}]. We numerically evaluate~\cite{Johansson+:12,Johansson+:13,Hunter:07,
Harris:20} the dynamics governed by the interaction Hamiltonian
\begin{equation} \label{eq:Ht} \begin{split} 
H_I(t) = {} & H^{\textrm{off}}_3(\vec{k}_0) \\ 
&+\sum_{j=1}^3\sum_{n=1}^3\left[\hbar\Omega_j(t)e^{i(\vec{k}_j\cdot\vec{x}_n
-\delta_j t)}\ket{r}_{nn}\bra{0} + \textrm{H.c.}\right] 
\end{split} \end{equation}
with initial state $\ket{\psi(t=0)}=\ket{ggg}$ and the time dependencies of $\Omega_j(t)$ 
chosen in the form of step functions, such that they equal the respective constant values 
at all times. Here we disregard the fast dynamics due to $H_A$ and the strong laser field with Rabi 
frequency $\Omega_0$ and detuning $\Delta_0$, because we are only interested in the slower dynamics 
introduced by the three fields $j=1,2,3$.

\begin{figure}
\includegraphics[width=\linewidth]{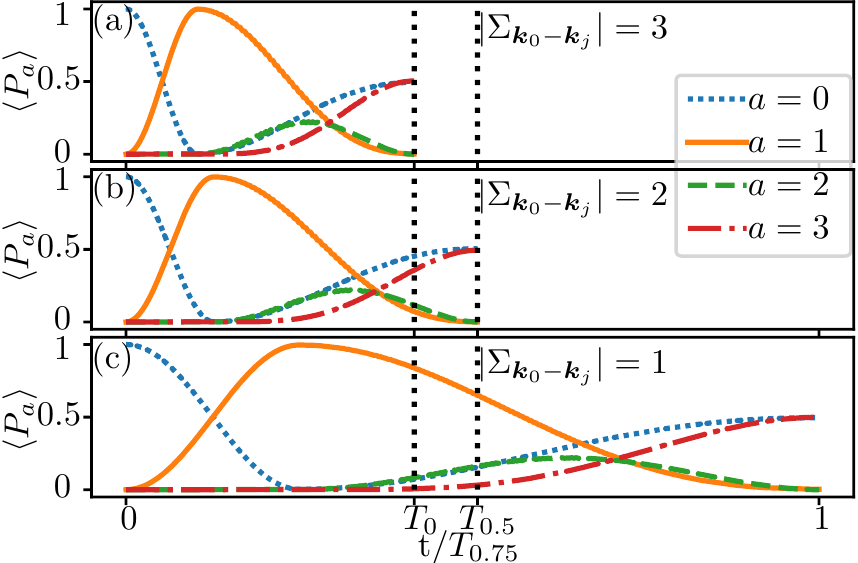}
\caption{\label{fig:conversionschemes_twisted} Expectation values $\langle P_a\rangle=\bra{\psi(t)}P_a\ket{\psi(t)}$ for 
the atomic excitation number $a$ over the conversion time for a conversion scheme $\ket{ggg}\stackrel{\pi-\text{pulse}}
{\longrightarrow} \ket{D^3_1(\vec{k}_0)}
\stackrel{[45]}{\longrightarrow}$ GHZ state for different amounts of relative twisting corresponding to 
$|\Sigma_{\vec{k}_0-\vec{k}_j}|=3,2,1$ (a-c). $T_{0,0.5,0.75}$ is the respective conversion time under the assumption
of equal laser-pulse energy consumption [cf. Eq.~\eqref{eq:laserenergy}].}
\end{figure}

For definiteness, we set $\Omega_0=-0.03\Delta_0$ and $\hbar\Delta_0/V = -0.7$.
The negative detuning $\Delta_0$ with respect to the Rydberg transition ensures that the field is even more detuned 
with respect to transitions involving the RB, hence $|s_0|>|s_1|>|s_2|$ with $T_0\:s_0/\hbar=-1247$ where $T_0$ in 
the considered conversion time in the case without relative twisting. The relation between detuning and the interaction 
energy shift $V$ ensures that all residual frequencies satisfy the condition $\min\{|\omega_R|\}T_0 > 600$.
The obtained numerical results are presented in Fig.~\ref{fig:conversionschemes_twisted}.
Similar to the atomic Hamiltonian $H_A$ which is not considered here, the first-order 
correction $H^\text{off}_3(\vec{k}_0)$ and the relative twisting give rise to phases [cf. Eqs.~\eqref{eq:statestwistedladder}]. Since we are not interested in such relative phases, we just show expectation 
values $\bra{\psi(t)}P_a\ket{\psi(t)}$.
As was to be expected, the population is transferred to the one-excitation subspace ($a=1$) via the $\pi$-pulse and the ensuing 
conversion scheme leads to a GHZ state, such that $\bra{\psi(T_k)}P_0\ket{\psi(T_k)}=\bra{\psi(T_k)}P_3 \ket{\psi(T_k)}=0.5$.
The required conversion time becomes significantly longer for a larger relative twisting, i.e. for smaller 
$|\Sigma_{\vec{k}_0-\vec{k}_j}|$.

With this approach by an additional laser field the reference frame of the twisted states participating in the effective dynamics 
can be set by the alignment of the strong field ($j=0$). Furthermore, even if -- for experimental reasons -- perfect alignment of the 
other fields or perfect positioning of the atoms is not possible, this can be partially compensated for by an adjustment of the Rabi 
frequencies. This speaks in favor of the flexibility of the proposed scheme.

To summarize, within the proposed state-conversion scheme suitable combinations of interatomic distances and laser 
orientations allow control of site-dependent phases. This opens up the possibility to address different states during 
the conversion process, with only a slight adjustment of the Rabi frequencies of external lasers.
It should be emphasized that, while a strong field is not necessary for our scheme, such a field can still be 
used for selecting -- in combination with fine detunings addressing lifted degeneracies -- twisted 
states that participate in the laser-controlled dynamics.
\subsection{Timescale hierarchies and conversion times} \label{TimeHierarch}
Various schemes for generating entanglement in quantum systems can be divided into those based on 
controlled dissipation~\cite{Rao+Moelmer:13,Carr+Saffman:13} and those that are governed by timescale hierarchies~\cite{Shankar+:13}.
Our state-conversion scheme in Sec.~\ref{ssec:twistcompensation} belongs to the latter group of schemes, which generically 
entail the application of a strong ``dressing drive'' at rate $G$ simultaneously with other interactions that are 
characterized by rates $g_i$. The linchpin of such schemes is that the dressing drive creates resonances that are 
resolved by the other drives in the limit $g_i \ll G$, and the corresponding hierarchy of timescales $g_i^{-1} \gg G^{-1}$ 
is what protects the entangled target state. 

It is important to stress that for all schemes based on timescale hierarchies the steady-state entanglement fidelity only 
asymptotically approaches unity upon increasing the relative strength 
$G/\textrm{max}\{g_i\}$ of the dressing drive. At the same time, timescale hierarchies limit the entanglement-generation 
speed, because the other interactions $g_i$ populating the entangled target state must be driven slowly compared to 
experimentally achievable rates for $G$.

To demonstrate the soundness of our proposed state-conversion scheme it is important to show that -- despite 
the limitations imposed by the aforementioned timescale hierarchies -- our characteristic state-conversion times 
are significantly shorter than the relevant Rydberg-state lifetimes. The latter scale as $\tau_n\propto n^3$, where 
$n$ is the principal quantum number, so that for $n\sim 50$ one has $\tau_n\sim 100\:\mu$s~\cite{GallagherBOOK}.
In particular, the hierarchy of timescales in the system at hand dictates the following inequalities for the pulse
duration $T_{\textrm{int}}$, the Rabi frequencies $g_i$, the spontaneous decay rate $\kappa/n^3$ of the $n$-th 
Rydberg state with $\kappa$ denoting a typical spontaneous decay rate of an energetically low-lying bound state, 
and the Stark-induced level shift $G$ (all expressed in frequency units):
\begin{equation}
\kappa/n^3 \ll T^{-1}_{\textrm{int}}\lesssim |g_i|\ll |G|\ll E_{\textrm I}/(\hbar n^3) \:.
\end{equation}
Here $E_{\textrm I}/(\hbar n^3)$ is the frequency corresponding to the level spacing between Rydberg states 
$n$ and $n+1$, with $E_{\textrm I}/\hbar = 10^{16}$\:s$^{-1}$ being its counterpart corresponding to the 
ionization energy $E_{\textrm I}\approx 13,6$\:eV of the hydrogen atom, and $\kappa = 10^9$ s$^{-1}$ (note 
that $\kappa$ is seven orders of magnitude smaller than $E_{\textrm I}/\hbar$ due to 
$\hbar \kappa/E_{\textrm I}$ being proportional to the third power of the fine-structure constant $\alpha\approx 1/137$).

The last conditions can be fulfilled, for example, for a Rydberg state with $n=50$ by choosing the relevant parameters 
such that $|G| = 10^{10}/125 s^{-1}$ and $|g_i| = 10^{8}/125 s^{-1}$. This further yields
\begin{equation}
125\times 10^{-10} \ll 125\times 10^{-8} s \lesssim T_{\textrm{int}} \ll 125 \times 10^{-6}s \:.
\end{equation}

Because the difference between spontaneous-decay rates and typical optical transition frequencies 
always involves a factor of $\alpha ^3\sim 10^{-7}$, these last conditions imply that the characteristic
state-conversion times in the system at hand are much smaller than the typical Rydberg-state lifetimes 
even if the rates $g_i$ and $G$ differ by a factor of $100$.
Therefore, typical conversion times in a neutral Rydberg-atom system are of the order of microseconds.

\section{Summary and Conclusions} \label{sec:conclusion}
To summarize, in this paper we addressed the problem of dynamical state generation (i.e. state preparation 
and conversion) in the Rydberg-blockade regime of a neutral-atom system in which the atomic ensemble is subject to multiple 
external laser fields. We presented a preparation scheme for twisted $W$ states, which 
assumes precise control over the alignment of resonant laser fields and the positioning of atoms.
We illustrated this scheme in the special case of three-qubit chiral $W$ states, a special type of $W$ 
states of importance for implementing noiseless-subsystem qubit encoding~\cite{Knill+:00}. 
In addition, we showed that conversions from twisted $W$- to GHZ states are possible 
by adopting already known pulse schemes for ordinary $W$ states.
We further showed, that even if such a precision in positioning of the atoms is not possible a state 
conversion starting from twisted $W$ states is still possible. It involves a downward renormalization 
of the relevant Rabi frequencies. Thus, somewhat higher laser-pulse energies are required to carry out the desired 
state conversion within the same time frame. We demonstrated the soundness of our state-conversion scheme 
by showing that the typical state-conversion times are much smaller than relevant Rydberg-state lifetimes.

Several possible directions of future work can be envisioned.
Firstly, while all the examples of state-engineering in the present work pertained to a three-atom (qubit) system, 
the preparation of general twisted $W$ states in lattice-periodic systems is of utmost importance in the area of analog 
quantum simulation~\cite{Georgescu+:14,Weimer+:21}.
Namely, owing to their known connection with single-excitation Bloch states, such states represent the desired states 
of analog simulators~\cite{Stojanovic+:12,Mei+:13,Stojanovic+:14,Stojanovic+Salom:19} prior to performing interaction 
quenches of various types~\cite{Hofer+:12}.
Secondly, while our proposed state-conversion scheme is characterized by timescale hierarchies, it would be instructive 
to devise its counterparts based on controlled dissipation~\cite{Rao+Moelmer:13,Carr+Saffman:13}.
Last but not least, the state preparation scheme proposed here can be extended to other classes of generalized $W$ states, 
different from the twisted ones. For instance, an interesting $W$-type state was proposed in the past for applications 
in quantum teleportation and superdense coding~\cite{Agrawal+Pati:06}.
However, this state has never been realized with Rydberg-atom-based qubits.  

\begin{acknowledgments}
The authors acknowledge useful discussions with G. Birkl. This research was supported by the Deutsche 
Forschungsgemeinschaft (DFG) -- SFB 1119 -- 236615297.
\end{acknowledgments}

%\newpage
\onecolumngrid
\appendix
\section{Derivation of the effective Hamiltonians}\label{app:effH}
\subsection{Off-resonant field} \label{app:effHoffresonant}
In what follows, we provide a detailed derivation of the effective Hamiltonian for a field (enumerated by $j=0$), 
which is far from resonance to any transition in $H_A$.

With the projectors $P^{m}_a$ discussed at the beginning of Sec.~\ref{ssec:offresonant}, we have $P_a^m H_\textrm{int}P_a^m=0$ 
and $\sum_{m=0}^\infty \sum_{a=0}^N P_a^m H P_a^m=H_0$, such that the first-order effective Hamiltonian is
\begin{equation}\label{eq:AppHeffoff}
H_{\textrm{eff}} = H_0 + \sum_{m=0}^{\infty}\sum_{a=0}^N P^m_a H_\textrm{int} Q_a^m \frac{1}{E_a-Q_a^m H_0 Q^m_a} Q^m_a 
H_\textrm{int} P^m_a \: ,
\end{equation}
where $Q_a^m= \mathds{1}-P_a^m $. Furthermore, since the interaction Hamiltonian $H_\textrm{int}$ connects only atomic subspaces 
which differ by one excitation, we obtain
\begin{equation}
P_a^m H_\textrm{int}Q_a^m = P_a^m H_\textrm{int} P_{a-1}^{m+1}+ P_a^m H_\textrm{int} P_{a+1}^{m-1}
\end{equation}
and we can write down the corresponding terms of the effective Hamiltonian in Eq.~\eqref{eq:AppHeffoff} as
\begin{equation}
P^m_a H_\textrm{int} Q_a^m \frac{1}{E_a-Q_a^m H_0 Q^m_a} Q^m_a H_\textrm{int} P^m_a=P_a^m H_\textrm{int} \left(\frac{P^{m+1}_{a-1}}
{E_a^m-E_{a-1}^{m+1}}+\frac{P^{m-1}_{a+1}}{E_a^m-E_{a+1}^{m-1}}\right)H_\textrm{int}P_a^m \: .
\end{equation}
The energy differences in the denominators are given by
\begin{align}
E_a^m-E_{a\mp 1}^{m \pm 1} 
&=\hbar(\mp\omega\pm\omega_A)+ V\left[\uber{a}{2}-\uber{a\mp 1}{2}\right]
 =\begin{cases} -\hbar\Delta_0+\uber{a-1}{1} V=-\hbar\Delta_0+(a-1)V \\
+\hbar\Delta_0-\uber{a}{1} V=\hbar \Delta_0-a V
\end{cases}
\end{align}
with $\Delta_0 = \omega-\omega_A$. Since $[P_a^m,U(k)\otimes \mathds{1}_F]=0$, we can compensate the site dependent phases 
via the unitary transformation $U(\vec{k}_0)$ and have to introduce them back into the equation at the end. By setting 
$P_{a<0}^m=P_{a>N}^m=P_a^{m<0}=0$ we can write down 
the general term 
\begin{equation} \begin{split}
\mathcal{C}_a^m(\vec{k}_0)= {} &U^\dagger(\vec{k}_0)P_a^m H_\textrm{int} P_{a\mp 1}^{m\pm 1} H_\textrm{int} P_a^m  U(\vec{k}_0)  \\
= {} & U^\dagger(\vec{k}_0)P_a^m H_\textrm{int}
\sum_{n_1<\ldots <n_{a\mp 1}}^N\ket{\{n_1, \ldots ,n_{a\mp 1}\}}\bra{\{n_1, \ldots ,n_{a\mp 1}\}}\otimes\ket{m\pm 1}\bra{m\pm 1}  \\
&\times \left(\sum_{n'=1}^N d^*_0\ket{g}_{n'n'}\bra{r}a^\dagger_0   +\textrm{H.c.}\right)P_a^m 
 \\
= {} &U^\dagger(\vec{k}_0)P_a^mH_\textrm{int}
\sum_{n'=1}^N \sum_{n_1<\ldots<n_{a\mp 1}}^N  \\
&\times \left[(1-\chi_{\{n_1,\ldots,n_{a\mp 1}\}}^{\{n'\}})d_0^*\sqrt{m\pm 1}\ket{\{n_1, \ldots ,n_{a\mp 1}\}}
\bra{\{n_1, \ldots ,n_{a\mp 1}\} \cup \{n'\}}\otimes \ket{m\pm 1}\bra{m\pm 1-1} \right.
 \\
&\phantom{\times \big[}\left. + \chi_{\{n_1,\ldots,n_{a\mp 1}\}}^{\{n'\}}d_0\sqrt{m\pm 1+1}\ket{\{n_1, \ldots ,n_{a\mp 1}\}}
\bra{\{n_1, \ldots ,n_{a\mp 1}\} \setminus \{n'\}}\otimes\ket{m\pm 1}\bra{m\pm 1+1}\right]P_a^m \: ,
\end{split} \end{equation}
where we used the characteristic function ($\chi^A_B=1$ if $A\subseteq B$ and $0$ otherwise) to encode the annihilation 
effect of the atomic rising and lowering operators.
Calculating the action of $H_\textrm{int}$ from the left results in four terms for each combination of ($(n_1,\ldots,n_a),n,n')$.

Since $P^m_a$ projects onto the subspace containing $m$ photons and $a$ atomic excitations only one term 
per case, i.e. per sign $\pm$, survives, resulting in
\begin{align}
\mathcal{C}_a^m(\vec{k}_0)=& 
\sum_{n_1<\ldots<n_{a\mp 1}}^N \left. 
\begin{cases}
|d_0|^2(m+1)\cdot\sum_{n=1}^N (1-\chi_{\{n_1,\ldots,n_{a- 1}\}}^{\{n\}})\ket{\{n_1, \ldots ,n_{a-1}\} \cup \{n\}}\\
\phantom{|\tilde\Omega|^2(m+1)\cdot}
\sum_{n'=1}^N (1-\chi_{\{n_1,\ldots,n_{a- 1}\}}^{\{n'\}})
\bra{\{n_1, \ldots ,n_{a-1}\} \cup \{n'\}}\\
\ \\
|d_0|^2m\cdot\sum_{n=1}^N \chi_{\{n_1,\ldots,n_{a+ 1}\}}^{\{n\}}\ket{\{n_1, \ldots ,n_{a+ 1}\}  \setminus \{n\}} \\
\phantom{|\tilde\Omega|^2m\cdot}
\sum_{n'=1}^N \chi_{\{n_1,\ldots,n_{a+ 1}\}}^{\{n'\}}\bra{\{n_1, \ldots ,n_{a+ 1}\} \setminus \{n'\}} 
\end{cases}\right\}\otimes \ket{m}\bra{m}
\label{eq:mplusminus1A} \: .
\end{align}
Reintroducing the site-dependent phases and dividing the resulting expression into diagonal elements, that commute 
with $U(\vec{k}_0)\otimes \mathds{1}_F$ and appear $a$ times, and off-diagonal ones which 
do not commute with $U(\vec{k}_0)\otimes \mathds{1}_F$ and appear just once, we obtain
\begin{align}
P_a^m H_\textrm{int}P_{a\mp 1}^{m\pm 1} H_\textrm{int} P_a^m 
=& \left. 
\begin{cases}
|d_0|^2(m+1) P_a\left[U(\vec{k}_0)\operatorname{Hd}_2(N)U^\dagger
(\vec{k}_0) + a\right]P_a \otimes \ket{m}\bra{m} \:, \\
\ \\
|d_0|^2m P_a\left[U(\vec{k}_0)\operatorname{Hd}_2(N)U^\dagger(\vec{k}_0)+
(N-a)\right]P_a \otimes \ket{m}\bra{m} \:.
\end{cases}\right. 
\end{align}
Here we used the operator $\operatorname{Hd}_2$, as defined in Sec.~\ref{ssec:offresonant}, to write down all off-diagonal 
elements. Thus, the effective Hamiltonian in this case reads
\begin{equation}
H_{\textrm{eff},N}^{\text{off}} = H_0 + \sum_{a=0}^{N}\sum_{m=0}^\infty |d_0|^2 P_a^m \left[m\frac{ U(\vec{k}_0)\operatorname{Hd}_2(N)
U^\dagger(\vec{k}_0)+N-a}{\hbar\Delta_0- a V}-(m+1)\frac{U(\vec{k}_0)\operatorname{Hd}_2(N) U^\dagger(\vec{k}_0)+a}{\hbar\Delta_0- 
(a-1) V}\right]P_a^m \: .
\end{equation}
Assuming a coherent state of high mean photon number $M_0$ and tracing over the field degrees of freedom we 
derive the atomic ensemble Hamiltonian of Eq.~\eqref{eq:effHoff} with Rabi frequency $\Omega_0^2=|d_0|^2 M_0/\hbar^2$. 

\subsection{Resonant field} \label{app:effHresonant}
In what follows, we present a detailed derivation of the effective Hamiltonian for a field which is resonant with 
one of the transitions in $H_A$.

As discussed in Sec.~\ref{ssec:resonant}, we have to join the two subspaces resonantly connected by the laser field.
Hence
\begin{equation}
P_{a, a-1}^{m, m+1} = P_a^m + P_{a-1}^{m+1} = \mathds{1}-Q_{a, a-1}^{m, m+1} \: , 
\end{equation}
which joins the terms containing $P_a^m$ and $P_{a-1}^{m+1}$ in the effective Hamiltonian.
For this resonance projector we have [cf. Eq.~\eqref{eq:1stHeff}]
\begin{equation}
P_{a, a-1}^{m,m+1}HP_{a, a-1}^{m,m+1} = P_{a, a-1}^{m,m+1}H_0 P_{a, a-1}^{m,m+1}
+ \left( P_a^m H_\textrm{int}P_{a-1}^{m+1} + \textrm{H.c.}\right) \: ,
\end{equation} 
where in comparison to the other non-resonance projectors an additional term appears, which contains $H_\textrm{int}$. 
Compensating for the site-dependent phases, this term can be written down as
\begin{equation} \begin{split}
\mathcal{C}_{a, a-1}^{m, m+1}= {} & U^\dagger(\vec{k_a})P_a^m H_\textrm{int} P_{a-1}^{m+1}U(\vec{k_a})  \\
= & \sum_{n_1<\ldots<n_a}^N \ket{\{n_1,\ldots,n_a\}}\sum_{n'_1<\ldots<n'_{a-1}}^N d_a\sqrt{m+1} 
 \\
&\times  \sum_{n=1}^N \chi^{\{n_1,\ldots,n_a\}}_{\{n_1',\ldots,n'_{a-1}\}\cup \{n\}}\left(1-\chi^{\{n\}}_{\{n_1',\ldots,n_{a-1}'\}}\right)
\bra{\{n_1',\ldots,n_{a-1}'\}} 
\otimes \ket{m}\bra{m+1} \Bigg. 
 \\
= {} &  d_a\sqrt{m+1}\sum_{n'_1<\ldots<n'_{a-1}}^N\sum_{n=1}^N \ket{r}_{nn}\braket{g}{\{n'_1,\ldots,n'_{a-1}\}} 
\bra{\{n'_1,\ldots,n'_{a-1}\}} \otimes \ket{m}\bra{m+1}  \\
= {} & d_a \sqrt{m+1} \cdot \sigma^+_{a-1} \otimes \ket{m}\bra{m+1} 
\end{split} \end{equation}
Here, we have introduced the rising operator of the atomic subspace with $a-1$ excitations, connecting this subspace 
to its counterpart with $a$ excitations. We can define a lowering operator in an equivalent fashion.
These two operators are given by 
\begin{equation} \begin{split}
\sigma_a^- &= \sum_{n_1<\ldots<n_a}^N \sum_{n=1}^N \ket{g}_{nn}\braket{r}{\{n_1,\ldots,n_a\}}\bra{\{n_1,\ldots,n_a\}} 
= \left(\sigma_{a-1}^+\right)^\dagger \:,\\
\sigma_a^+ &= \sum_{n_1<\ldots<n_a}^N \sum_{n=1}^N \ket{r}_{nn}\braket{g}{\{n_1,\ldots,n_a\}}\bra{\{n_1,\ldots,n_a\}} 
= \left(\sigma_{a+1}^-\right)^\dagger \:,
\end{split} \end{equation}
and act on the states $\ket{D^N_a}$ according to
\begin{equation} \begin{split}
\sigma_a^+ \ket{D^N_a} &= \sqrt{(N-a)(a+1)}\ket{D^N_{a+1}} \:,\\
\sigma_a^- \ket{D^N_a} &= \sqrt{a(N-a+1)}\ket{D^N_{a-1}} \:.
\end{split} \end{equation}
In addition, for the second part of the effective Hamiltonian we have to compute
\begin{equation} \begin{split}
P_{a,a-1}^{m,m+1}HQ_{a,a-1}^{m,m+1}&\frac{1}{E_a^{m}-Q_{a, a-1}^{m,m+1}HQ_{a, a-1}^{m,m+1}}
Q_{a,a-1}^{m,m+1}HP_{a,a-1}^{m,m+1}  \\
&=P_{a,a-1}^{m,m+1}H_\textrm{int}Q_{a,a-1}^{m,m+1}\frac{1}{E_a^{m}-Q_{a, a-1}^{m,m+1}H_0Q_{a, a-1}^{m,m+1}}
Q_{a,a-1}^{m,m+1}H_\textrm{int}P_{a,a-1}^{m,m+1}  \\
&= P_a^m H_\textrm{int} \frac{P_{a+1}^{m-1}}{E^m_a-E^{m-1}_{a+1}}H_\textrm{int}P_a^m+P_{a-1}^{m+1} H_\textrm{int} 
\frac{P_{a-2}^{m+2}}{E^{m+1}_{a-1}-E^{m+2}_{a-2}}H_\textrm{int}P_{a-1}^{m+1} \:.
\end{split} \end{equation}

With an index shift $a \to a-1$ and $m\to m+1$ to match the second term we can use the results pertaining 
to the off-resonant field in Appendix~\ref{app:effHoffresonant}.
For all other projectors with $a'\neq a, a-1$ we can also use the results corresponding to the off-resonant
case, because for other transitions the resonance condition is not fulfilled. Putting everything together yields
\begin{align}
U^\dagger(\vec{k_a})H_{\textrm{eff},N}^{a\leftrightarrow a-1}U(\vec{k_a}) = H_0  + \sum_{m=0}^\infty
\Bigg[& \left(d^* \sqrt{m+1} \cdot \sigma^-_a \otimes \ket{m+1}\bra{m} + \textrm{H.c.}\right)
- \frac{|d_a|^2 m}{V} \left(\operatorname{Hd}_2(N)
+(N-a)\right)P_a^m 
\nonumber \\
& -\frac{|d_a|^2(m+2)}{V}\left(
\operatorname{Hd}_2(N)
+a-1\right)P_{a-1}^{m+1}
 \nonumber  \\
&+\sum_{a,a-1\neq a'=0}^{N} |d_a|^2 \left(m\frac{
	\operatorname{Hd}_2(N)
	+(N-a')}{\hbar\Delta_a-a'V}-(m+1)\frac{
	\operatorname{Hd}_2(N)
	+a'}{\hbar\Delta_a-(a'-1)V}\right)P_{a'}^m\Bigg. \Bigg]  \: .
\end{align}
By further assuming that $V\gg\hbar|\Omega_a|$ for all fields, we can ignore all terms scaling 
with $|d_a|^2/V$ and only the first line contributes. Tracing out the field degrees of freedom 
and assuming a coherent field state results in the effective Hamiltonian as given by Eq.~\eqref{eq:effHres}.

\section{Detailed derivation of the twisted-ladder Hamiltonian} \label{app:twistcompensation}
In the following, we provide detailed derivation of the twisted-ladder Hamiltonian in 
Eq.~\eqref{eq:twistedLadder} starting from Eq.~\eqref{eq:ladderHoff}.

We can evaluate the three terms ($a=1,2,3$) of Eq.~\eqref{eq:ladderHoff} separately using the 
well-known operator identity
\begin{equation}
e^{A}Be^{-A}=\sum_{m=0}^\infty\frac{1}{m!}\left[A,B\right]_m \:,
\end{equation}
where $\left[A,B\right]_m$ is a shorthand for the repeated commutator of $A$ and $B$ with $m$ appearances 
of the operator $A$. This is relatively straightforward for $a=1,2,3$. For the sake of brevity, we omit the 
argument $\vec{k}_a-\vec{k}_0$ of $U(\vec{k}_a-\vec{k}_0)$, because in each $a$-term the argument is the 
same. We reinstate this argument at the end of the derivation. 

We first evaluate
\begin{align}
[H^\text{off},U\sigma_1^-U^\dagger]&=(3s_0-s_1) U\sigma_1^-U^\dagger+(s_0-s_1)\Sigma_{\vec{k}_0-\vec{k}_1}\sigma_1^-
\end{align}
where $s_a := \hbar^2|\Omega_0|^2/(\hbar\Delta_0-aV)$ are the energy shifts introduced by $H^\text{off}_3$ and 
$H^\text{off}=H^\text{off}_3(\vec{k}=0)$.
We can see that the commutator partially reproduces the operator and adds an additional term without twisting.
Therefore, we can write down the $m$-th commutator using a triangular matrix as
\begin{align} \label{eq:effHrotated}
e^{iH^\text{off} t/\hbar}U\sigma_1^- U^\dagger e^{-iH^\text{off}t/\hbar}=\sum_{m=0}^\infty\frac{(i t/\hbar)^m}{m!}
\left[H^\text{off},U\sigma_1^-U^\dagger\right]_m = \sum_{m=0}^\infty \frac{(i t/\hbar)}{m!}  \vec{v}_1 \cdot \left(A_1^m\vec{e}_1
\right)=\vec{v}_1 \cdot \left(e^{it A_1/\hbar}\vec{e}_1\right)
\end{align}
with $\vec{e}_1=\begin{pmatrix}1& 0 \end{pmatrix}^T$ being a unit vector of appropriate dimension and
\begin{equation}
\vec{v}_1=\begin{pmatrix} U\sigma_1^-U^\dagger \\ \sigma_1^- \end{pmatrix} \ ; \
\ A_1= \begin{pmatrix} 3s_0-s_1 & 0 \\ (s_0-s_1)\Sigma_{\vec{k}_0-\vec{k}_1} & 6s_0-4s_1 \end{pmatrix} \ 
\end{equation}
with eigenvalues $\sigma(A_1)=\{3s_0-s_1,6s_0-4s_1\}$. Similarly, for $a=3$ we obtain
\begin{equation}
[H^\text{off},U\sigma_3^-U^\dagger]=(-s_1+3s_2) U\sigma_3^-U^\dagger+\sigma_3^-(-s_1+s_2)\Sigma_{\vec{k}_0-\vec{k}_3} \ 
\end{equation}
and the analogous expression with a matrix exponential and
\begin{equation}
\vec{v}_3=\begin{pmatrix} U\sigma_3^-U^\dagger \\ \sigma_3^- \end{pmatrix} \ \ ; \ 
\ A_3=\begin{pmatrix} -s_1+3s_2 & 0 \\ (-s_1+s_2)\Sigma_{\vec{k}_0-\vec{k}_3} & -4s_1+6s_2 \end{pmatrix} \ 
\end{equation}
with eigenvalues $\sigma(A_3)=\{-s_1+3s_2,-4s_1+6s_2\}$.
The case $a=2$ is more complicated and we will compute the commutator for a more general case involving 
the unitary transformations $U'=U(\vec{k}')$ and $U^\dagger=U^\dagger(\vec{k})$ for different wave vectors. 
This will be helpful later on for defining the matrix $A_2$. We calculate
\begin{equation} \begin{split}
\left[H^\text{off},U'\sigma_2^-U^\dagger\right]= {} & \left[H^\text{off},3 \ket{D^3_1(\vec{k}')}\bra{D^3_2(\vec{k})}
-\Phi(-\vec{k})\sum_{n=1}^3e^{i(\vec{k}'+\vec{k})\cdot\vec{x}_n}\ket{gg}\ket{r}_{nn}\bra{g}
\bra{rr}\right]  \\
= {} & 2s_1 U'\sigma_2^-U^\dagger  \\
&+3\Sigma_{\vec{k}'} (-s_0+s_1)\ket{D^3_1(0)}\bra{D^3_2(\vec{k})}
+3 \Sigma_{\vec{k}}(s_1-s_2)
\Phi(-\vec{k})\ket{D^3_1(\vec{k}')}\bra{D^3_2(0)}  \\
&+3\Phi(\vec{k}')(s_0-s_1)\ket{D^3_1(0)}\bra{D^3_2(\vec{k}'+\vec{k})}
+3\Phi(-\vec{k})(-s_1+s_2)\ket{D^3_1(\vec{k}'+\vec{k})}\bra{D^3_2(0)} \: , 
\end{split} \end{equation}
with $\Phi(\vec{k})=e^{i\vec{k}\cdot\sum_n^N \vec{x}_n}$.
Since all but the self reproducing part (first line on the right hand side) contains at least one generalized 
$D^3_{1,2}$-state (either ket or bra) with $\vec{k}=0$ the next commutator will accumulate terms of the form 
$\ket{D^3_1(0)}\bra{D^3_2(0)}$.
Omitting the argument $\vec{k}=0$ in the following we can write down the transformed $a=2$ part similar as the 
other ones by using $\vec{k}'=\vec{k}$.
Resulting in an analogues matrix exponential equation with
\begin{equation} \begin{gathered}
\vec{v}_2=\begin{pmatrix} U\sigma_2^-U^\dagger \\ \ket{D^3_1}\bra{D^3_2(\vec{k})} \\ \ket{D^3_1(\vec{k})}
\bra{D^3_2}\\ \ket{D^3_1}\bra{D^3_2(2\vec{k})} \\ \ket{D^3_1(2\vec{k})}\bra{D^3_2} \\ \ket{D^3_1}\bra{D^3_2} \end{pmatrix}
\ \ ; \ \ \vec{e}_1 = \begin{pmatrix} 1 \\ 0 \\ 0 \\ 0\\ 0 \\ 0 \end{pmatrix} \ ;
\\
A_2= \footnotesize
\left( \begin{array}{cccccc}
2 s_1 & 0 & 0 & 0 & 0 & 0 \\
3\Sigma_{\vec{k}}(-s_0+s_1) & -3s_0+5s_1 & 0 & 0 & 0 & 0 \\
3 \Phi(-\vec{k})\Sigma_{\vec{k}}(s_1-s_2) &0 & 5s_1-3s_2 & 0 & 0 & \\
3\Phi(\vec{k}) (s_0-s_1) & 0 & 0 & -3s_0 +5s_1 & 0 & 0 \\
3\Phi(-\vec{k})(-s_1+s_2) & 0 & 0 & 0 & 5s_1-3s_2 & 0 \\
0 & \Phi(-\vec{k}) \Sigma_{\vec{k}}(s_1-s_2) & \Sigma_{\vec{k}}(-s_0+s_1) & \Phi(-2\vec{k})
\Sigma_{2\vec{k}}(s_1-s_2) & 
\Sigma_{2\vec{k}} (-s_0 + s_1) & -3s_0 + 8s_1 - 3s_2 
\end{array}\right) \ 
\end{gathered} \end{equation}
and substitute $\vec{k}=\vec{k}_2-\vec{k}_0$. The corresponding eigenvalues are $\{2s_1,-3s_0+5s_1,5s_1-3s_2,-3s_0+8s_1-3s_2\}$.
Given that in the three equations for $a=1,2,3$ only the transformed unit vectors are of interest, 
we can write the transformed ladder Hamiltonian [cf. Eq.~\eqref{eq:ladderHoff}] in the form
\begin{equation} \begin{split}
U(\vec{k}_0)e^{iH^\text{off} t/\hbar}U^\dagger(\vec{k}_0) & H^\text{L}_3(\{\vec{k}_a\})U(\vec{k}_0) e^{-iH^\text{off}t/\hbar}
U^\dagger(\vec{k}_0) =U(\vec{k}_0)\left[\sum_{a=1}^3\Omega^*_a\vec{v}_a \left(e^{it A_a/\hbar}
\vec{e}_1\right) +\textrm{H.c.}\right]U^\dagger(\vec{k}_0)  \\
= {} \Bigg[&\sqrt{3}\Omega^*_1\begin{pmatrix} \ket{ggg}\bra{D^3_1(\vec{k}_1)}   \\ \ket{ggg}\bra{D^3_1(\vec{k}_0)} 
\end{pmatrix} \cdot \begin{pmatrix} e^{it(3s_0-s_1)/\hbar} \\ (e^{it(6s_0-4 s_1)/\hbar}- e^{it(3s_0-s_1)/\hbar}) 
\Sigma_{\vec{k}_0-\vec{k}_1}/3 \end{pmatrix}  \\
& + \sqrt{3}\Omega^*_3\begin{pmatrix} \ket{D^3_2(\vec{k}_3)}\bra{rrr}\Phi(-\vec{k}_0) \\ \ket{D^3_2(\vec{k}_0)}\bra{rrr}
\Phi(-\vec{k}_0) \end{pmatrix} \cdot
\begin{pmatrix} e^{it(-s_1+3s_2)/\hbar}  \\ \left(e^{it(-4s_1+6s_2)/\hbar}-e^{it(-s_1+3s_2)/\hbar} \right) \Sigma_{\vec{k}_0
-\vec{k}_3}/3  \end{pmatrix}\\
&+\Omega^*_2 \begin{pmatrix}3\ket{D^3_1(\vec{k}_2)}\bra{D^3_2(\vec{k}_2)} \\ \ket{D^3_1(\vec{k}_0)}\bra{D^3_2(\vec{k}_2)} \\ 
\ket{D^3_1(\vec{k}_2)}\bra{D^3_2(\vec{k}_0)}\\ \ket{D^3_1(\vec{k}_0)}\bra{D^3_2(2\vec{k}_2-\vec{k}_0)} \\ \ket{D^3_1(2\vec{k}_2-
\vec{k}_0)}\bra{D^3_2(\vec{k}_0)} \\ \ket{D^3_1(\vec{k}_0)}\bra{D^3_2(\vec{k}_0)} \end{pmatrix} \cdot
\begin{pmatrix} e^{it2s_1/\hbar} \\  \left(e^{it(-3s_0+5s_1)/\hbar}-e^{it2s_1/\hbar}\right) \Sigma_{\vec{k}_2-\vec{k}_0} \\  
\left(e^{it(5s_1-3s_2)/\hbar}-e^{it2s_1/\hbar}\right) \Sigma_{\vec{k}_2-\vec{k}_0}\Phi(\vec{k}_0-\vec{k}_2)) \\ 
\left(-e^{it(-3s_0+5s_1)/\hbar}+e^{it2s_1/\hbar}\right) \Phi (\vec{k}_2-\vec{k}_0)\\ \left(e^{it2s_1/\hbar}-e^{it(5s_1-3s_2)/\hbar}
\right) \Phi (\vec{k}_0-\vec{k}_2) \\ \eta(\vec{k}_2-\vec{k}_0)\end{pmatrix}  \\
&-\Omega^*_2e^{it2s_1/\hbar}\Phi(-\vec{k}_2)\sum_{n=1}^3e^{i2\vec{k}_2\cdot\vec{x}_n}\ket{gg}\ket{r}_{nn}
\bra{g}\bra{rr}\Bigg]+\textrm{H.c.}
\end{split} \end{equation}
with
\begin{equation}
\eta(\vec{k}) =
\frac{2\Sigma_{-\vec{k}}}{3}\left[e^{it2 s_1/\hbar}-e^{it(-3 s_0+5 s_1)/\hbar}
- e^{it(5 s_1-3 s_2)/\hbar}+e^{it(-3s_0+8 s_1-3s_2)/\hbar}\right] \: .
\end{equation}
From the last equation we can identify one relevant term per field ($j=1,2,3)$ and compensate 
the exponential time dependence via fine detunings. Choosing
\begin{align}
\delta_1 = (-6s_0+4s_1)/\hbar \:,&& \delta_2 = (3s_0-8s_1+3s_2)/\hbar \:,  && \delta_3 = (4s_1-6s_2)/\hbar \:,
\end{align}
results in the effective Hamiltonian in which terms oscillating with non-vanishing residual 
frequencies are neglected (cf. Sec.~\ref{ssec:twistcompensation}). The set of residual frequencies $\{\omega_R\}$
is given by
\begin{align}
-3s_0+s_1-\hbar\delta_1 &= 3s_0 - 3s_1 \:, \nonumber \\
-2s_1 - \hbar\delta_2 = -3s_0 + 6s_1 - 3s_2  \: , \
3s_0-5s_1 -\hbar\delta_2 &= 3s_1-3s_2 \: , \
-5s_1+3s_2 -\hbar\delta_2 = -3s_0+3s_2 \: , \label{eq:resfreq2} \\
s_1-3s_2 - \hbar\delta_3 &= -3s_1+3s_2  \:, \nonumber 
\end{align}
where each equation corresponds to one residual energy $\hbar\omega_R$ and each line 
corresponds to one value of $a$ ($a=1,2,3$).

\twocolumngrid

\end{document}